\documentclass[aps,prapplied,reprint, superscriptaddress, longbibliography]{revtex4-2}

\usepackage[version=3]{mhchem} % Formula subscripts using \ce{}
\usepackage{xcolor} % added for editing purposes. See newcommands section.
\usepackage[T1]{fontenc}
\usepackage{physics}
\usepackage{graphicx}
\usepackage{cleveref}

\usepackage{xr}
\newcommand{\LCO}{\ce{LaCrO_3}~}
\newcommand{\crocr}{\ce{Cr^{3+} - O - Cr^{3+}}}

\begin{document}

\author{Carson O. Patterson}
\email{carpat4@uw.edu}
\thanks{contributed equally to this work.}
\affiliation{Department of Physics, University of Washington, Seattle, WA 98105, USA}
\author{Ethan Q. Williams}
\email{eqwilliams@fortlewis.edu}
\thanks{contributed equally to this work.}
\affiliation{Department of Physics, University of Washington, Seattle, WA 98105, USA}
\author{Eden Tzanetopoulos}
\affiliation{Department of Chemistry, University of Washington, Seattle, WA, USA}
\author{Benjamin C. Li}
\affiliation{Department of Physics, University of Washington, Seattle, WA 98105, USA}
\author{Rachel T. Smith}
\affiliation{Department of Chemistry, University of Washington, Seattle, WA, USA}
\author{Matthew Chang}
\affiliation{Department of Chemistry, University of Washington, Seattle, WA, USA}
\author{Hideyuki Watanabe}
\affiliation{Global Research and Development Center for Business by Quantum-AI Technology (G-QuAT), National Institute of Advanced Industrial Science and Technology (AIST), Tsukuba Central 2, 1-1-1 Umezono, Tsukuba, Ibaraki 305-8568, Japan}
\author{Daniel R. Gamelin}
\affiliation{Department of Chemistry, University of Washington, Seattle, WA, USA}
\author{Kai-Mei C. Fu}
\affiliation{Department of Physics, University of Washington, Seattle, WA 98105, USA}
\affiliation{Department of Electrical and Computer Engineering, University of Washington, Seattle, WA, USA}
\affiliation{Physical Sciences Division, Pacific Northwest National Laboratory, Richland, Washington 99352, USA}

\title{Single-particle detection of the Néel transition in \LCO microcrystals via widefield nitrogen-vacancy center magnetic imaging}

\begin{abstract}
Rare-earth orthochromites exhibit rich, chemically tunable magnetic behavior, making them attractive for temperature sensing and temperature-dependent switching applications. Realizing this potential requires understanding how synthesis conditions and particle morphology influence the underlying \ce{Cr^{3+}} magnetic ordering. Orthochromite magnetism has traditionally been characterized by ensemble techniques such as neutron diffraction and SQUID magnetometry, which report only averaged magnetic properties and cannot access  particle-level information. Here we use widefield nitrogen-vacancy (NV) magnetometry to image the stray magnetic fields of individual \LCO microcrystals produced via molten salt synthesis, resolving the antiferromagnetic Néel transition in single particles. Of the 18 particles measured between 278 and 298\,K, five show a clear onset of magnetization near the ensemble Néel temperature, four show no measurable magnetic signal, and nine display temperature-independent signatures consistent with localized magnetic impurities undetected by ensemble measurements. Below the N\'eel temperature, particle magnetization remains fixed under a rotating in-plane field, consistent with canted antiferromagnetic ordering rather than paramagnetism, while repeated thermal cycling shows that the magnetic field pattern varies between cooldowns with certain features recurring at the same locations, consistent with micron-scale structurally pinned antiferromagnetic domains in which the sign of the weak ferromagnetic moment is set independently each time they pass through the Néel transition. These results establish single-particle NV magnetometry as a sensitive probe of magnetic ordering and domain structure in individual microcrystals, a key step toward correlating individual particle morphology and defect structure with magnetic ordering in orthochromites.
\end{abstract}

\maketitle

\section{Introduction}

Rare-earth orthochromites ($R\ce{CrO3}$) exhibit rich, chemically tunable magnetic behavior with potential applications in temperature-dependent sensing and switching. Central to these applications is the \ce{Cr^{3+}} paramagnetic-to-antiferromagnetic Néel transition that can occur over a broad temperature range depending on the rare-earth ion and composition. Micro- and nanoscale particles are especially interesting for temperature sensing in settings inaccessible to bulk sensors, such as within microelectronic devices, microfluidic channels, or individual cells \cite{dacaninfar_luminescence_2023,cheng_precise_2020}. In these settings, the sensor is a single particle, making it critical to characterize the particle's transition rather than the ensemble-averaged transition or ensemble net magnetization. Any particle-to-particle spread in the Néel temperature ($T_N$), or any particle whose magnetic response is dominated by impurities rather than Cr$^{3+}$ order, limits the accuracy such a probe can reach. 

\LCO is a model orthochromite for studying this transition because its Néel temperature is near room temperature~\cite{jonker_magnetic_1956,Qahtan2024RareEarthChromitesReview}. Below $T_N$, the structural anisotropy of the tilted \ce{Cr^{3+}} octahedra gives rise to a Dzyaloshinskii-Moriya interaction that slightly cants all the chromium spins in a common direction, producing a net weak ferromagnetic moment along the $c$-axis \cite{zhou_intrinsic_2010, zhou_magnetic_2011, tiwari_magnetostructural_2015, silva_structural_2016, aamir_ferroelectric_2021, lahlou_nabil_morphological_2022, duran_lacro3_2025}. The spin ordering and magnetic phase transition of \LCO have been characterized primarily by neutron-diffraction experiments \cite{koehler_neutron-diffraction_1957, zhou_intrinsic_2010, zhou_magnetic_2011, gupta_spinphonon_2020} and superconducting quantum interference device (SQUID) magnetometry \cite{tiwari_magnetostructural_2015, barrozo_ferromagnetism_2013} on powder samples. Both techniques report only the average magnetic properties of an ensemble. Because the weak ferromagnetic moment arises directly from the crystallographic tilt of the \ce{CrO6} octahedra, the domain structure is expected to be sensitive to crystal morphology, faceting, and defect density, properties that vary between synthesis methods and between individual particles. Single-particle characterization is therefore needed to reveal how these factors influence orthochromite magnetic properties. 

Imaging antiferromagnetic order is generally difficult because the compensated spin sublattices produce no net moment outside the sample. Canted antiferromagnets are an exception. The Dzyaloshinskii-Moriya moment is rigidly tied to the staggered magnetization, so it both tracks the order parameter and generates a stray field that can be measured outside the particle; reversing the antiferromagnetic order parameter reverses the canted moment, and thus the sign of the stray field. Established probes of antiferromagnetic domains, such as X-ray magnetic linear dichroism photoemission electron microscopy~\cite{folven_effects_2011}, are well suited to flat, extended films but require stringent surface and vacuum conditions and, being quadratic in the order parameter, are insensitive to its sign. Magnetometry with nitrogen-vacancy (NV) centers is well suited to imaging canted antiferromagnetic order in individual microparticles because it can map both the magnitude and direction of microtesla-scale magnetic fields to sub-micron resolution, requires no sample preparation beyond deposition onto the diamond surface, and reports both the local magnetic field and the local temperature.

NV magnetometry has been applied to magnetic ordering and to individual particles, but largely in separate contexts. In extended thin films, where sample geometry supports both scanning probes and surface-sensitive techniques, NV magnetometry has resolved antiferromagnetic domains 
\cite{Li2023Mn3SnNVImaging, appel_nanomagnetism_2019, gross2017rsi}, imaged domain walls \cite{tetienne_nanoscale_2014, tetienne_nature_2015} and ferroelectric domains \cite{huxter_imaging_2023}, and mapped ferromagnetic domain structure~\cite{simpson_magneto-optical_2016}. In particulate systems, NV magnetometry has probed Néel relaxation in superparamagnetic particles \cite{richards_time-resolved_2025}, clustering of nanoproteins \cite{lamichhane_nitrogen-vacancy_2024}, and the magnetic character of individual iron-triazole spin-crossover nanorods~\cite{lamichhane_nitrogen-vacancy_2023}. These particulate studies characterize the magnetic state of a particle at a fixed temperature or its response to an applied field rather than following a particle through its own magnetic ordering transition. 

In this work, we use widefield NV magnetometry to image the stray fields of eighteen individual LaCrO$_3$ microcrystals. We observe a variety of magnetic responses across this set of microcrystals, ranging from the clear onset of magnetization near the Néel temperature to no measurable magnetic signal. For the transitioning particles, the magnetization remains fixed as an in-plane applied field is rotated, as expected for antiferromagnetic ordering rather than paramagnetism. Repeated thermal cycling through $T_N$ produces field patterns that vary from cooldown to cooldown but contain recurring features, consistent with a small number of magnetic domains whose positions are pinned by structural features while the sign of each domain's moment is set stochastically at each cooldown. This single-particle approach reveals magnetostructural detail that ensemble techniques cannot access, opening a path toward understanding how morphology governs magnetic ordering in orthochromite microparticles.

\section{Canted antiferromagnetism and domain structure in L\lowercase{a}C\lowercase{r}O$_3$}\label{sec:model}

The dominant source of the measured magnetization in \LCO is the spin of the \ce{Cr^{3+}} ions. Above the Néel temperature, the \ce{Cr^{3+}} spins are in a paramagnetic state with no long-range ordering (Fig. \ref{fig:master_plot_all_particles}a). Below the Néel temperature, \LCO exhibits G-type antiferromagnetism in which each \ce{Cr^{3+}} spin is antiparallel to all six nearest neighbors, forming two oppositely polarized spin sublattices \cite{kittel_introduction_1996, koehler_neutron-diffraction_1957}. The \ce{CrO6} octahedra in the \textit{Pbnm} structure are tilted by 
$\sim$\,10$^\circ$, resulting in a \crocr\ bond angle of $\sim$160$^\circ$ \cite{zhou_magnetic_2011}. The bend removes the inversion center at the midpoint of the Cr-O-Cr bond, which, in the presence of spin-orbit coupling, permits an antisymmetric Dzyaloshinskii-Moriya interaction \cite{dzyaloshinsky_thermodynamic_1958, moriya_anisotropic_1960} that cants spins on both sublattices in a common direction. The canting angle $\beta$ between spins on the two sublattices is expected to be small, $\beta \sim$\,$0.5^\circ$ \cite{kim_effect_2011}, producing a small net ferromagnetic moment aligned along the $c$-axis (Fig. \ref{fig:master_plot_all_particles}a) \cite{zhou_magnetic_2011}. This paramagnetic-to-antiferromagnetic transition in \LCO is a second-order phase transition with the staggered magnetization $M_{stag} = \langle M_A - M_B \rangle/2$ as the order parameter, where $M_A$ and $M_B$ are the magnetizations of the two spin sublattices \cite{goldenfeld_lectures_2018}. Above $T_N$, $M_{stag} = 0$; below $T_N$, spontaneous symmetry breaking selects a nonzero value of $M_{stag}$. The stray field beneath a \LCO particle measured in our experiment is determined by the net canted moment $M_\mathrm{net} = 2M_{stag}\sin(\beta/2)$ and therefore proportional to $M_{stag}$. 

Multiple antiferromagnetic domains distinguished by easy-axis orientation have been directly observed in thin films of \ce{LaFeO3} \cite{folven_effects_2011}, which shares the same orthorhombic structure and canted antiferromagnetism as \LCO. In a canted antiferromagnet the magnetostatic energy is too small to drive domain formation at our particle sizes. Taking the magnetocrystalline anisotropy energy density $K \sim 10^4$\,J\,m$^{-3}$, consistent with antiferromagnetic resonance measurements in the orthochromite YCrO$_3$~\cite{ikeda2015hfe}, and $M_\mathrm{net}\approx4\times10^3$\,A m$^{-1}$ from the canting angle above, the single domain critical radius is several hundred microns~\cite{kittel_physical_1949}, an order of magnitude larger than the particles studied in this work. The domain size is instead set by the scale of the structural domains inherited from crystal growth. The correlation length surviving the ordering transition can, in principle, provide a second bound, but the relevant spin dynamics are orders of magnitude faster than the thermal ramps used here~\cite{kimel2004liu}, so the structural scale is expected to dominate. For LaCrO$_3$, the tilt pattern is directly coupled to the Dzyaloshinskii-Moriya vector and the easy axis~\cite{zhou_magnetic_2011}; since different structural twin variants correspond to different orientations of the tilt pattern, we expect a magnetic domain wall to coincide with each twin boundary.

\section{Experimental}

\subsection{\LCO microcrystals}\label{sec:lacro3_synthesis}
1\% ytterbium-doped \LCO microcrystals with characteristic lengths of 10-50\,\textmu m (Fig.~\ref{fig:experimentalschematic}) were synthesized by adapting a molten salt synthesis method described in~\cite{tzanetopoulos_luminescent_2026}. A 0.5 molar salt mixture was made with ratios of 1:2 lanthanide:chromium and 1:1 sodium chloride:potassium chloride (both $\geq$99.0\%, Sigma). The lanthanide component was composed of 99\% lanthanum nitrate hexahydrate (99.99\%, Sigma) and 1\% ytterbium nitrate pentahydrate (99.99\%, Sigma). Chromium nitrate nonahydrate (99\%, Sigma) was used as the chromium source. The synthesis procedure is described in detail in the Supplemental Material Sec. 1 \cite{supplement}. 

1\% ytterbium was added to enable photoluminescence (PL) measurements, which can detect the Néel transition through the resulting exchange interaction in orthochromites, as described by Tzanetopoulos \emph{et al.}~\cite{tzanetopoulos_luminescent_2026}. At this doping level, ytterbium is not expected to affect \LCO's magnetic properties. Consistent with this, zero-field-cooled SQUID susceptibility on an ensemble of these microcrystals gives $T_N$ = 287\,K (Supplemental Material Sec. 5), within the 287–295\,K range reported for undoped bulk LaCrO$_3$ from magnetization measurements~\cite{tiwari_magnetostructural_2015,silva_structural_2016,barrozo_ferromagnetism_2013}. However, the ytterbium PL lines in this sample were too broad to resolve the expected exchange splitting at any temperature between the Néel transition and 4\,K (Fig. S6 ~\cite{supplement}). This result further motivates single-particle magnetic characterization via NV magnetic imaging.

\subsection{Magnetic Particle Imaging}
\label{subsec:magpi}

\begin{figure*}[t]
 \includegraphics[width=0.9\linewidth]{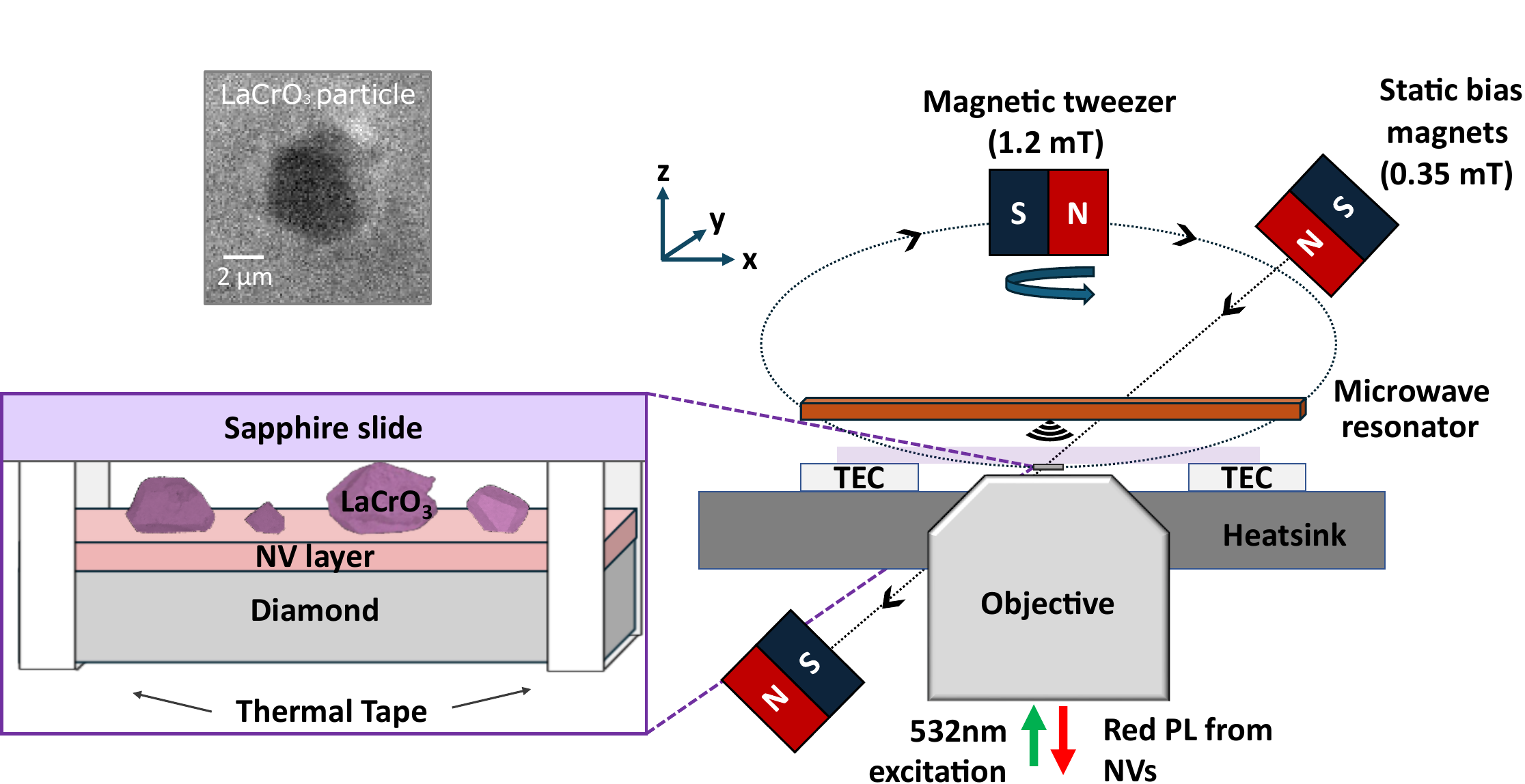}
\caption{The NV diamond microscope. \LCO samples were dropcast onto the diamond surface and the diamond was taped to a thermally conductive sapphire slide via thermal tape. The particles were imaged from below through the $\sim$100-\textmu m thick diamond sample. The sapphire slide was mounted to the sample stage, which also served as a heatsink, via two thermo-electric coolers (TECs). Permanent magnets provided a static bias field in the x-z plane while the magnetic tweezer added a rotating $1.2$\,mT bias field in the x-y plane. The inset shows a brightfield image of Particle 1 on the diamond sensor surface.}
\label{fig:experimentalschematic}
\end{figure*}

We performed widefield imaging of an NV ensemble to generate maps of the magnetic field produced by single particles (Fig. \ref{fig:experimentalschematic}). The diamond sensor consisted of a 150 nm $^{15}\text{N}$-doped layer grown on an electronic-grade diamond substrate with source gas that had a $^{12}\text{C}$ isotopic purity of $>99.99$\%. Vacancies were introduced by He$^+$ implantation followed by vacuum and O$_2$ anneals for NV formation and charge-state stabilization (Supplemental Material Sec. 2.1 \cite{supplement}). The NV density was $\sim$\,$1.7\times10^{16}$\,cm$^{-3}$ \cite{Kazi2021}. We imaged the NV photoluminescence onto a CMOS camera while sweeping the microwave frequency across the NV ground-state spin transitions, $\ket{m_s = 0} \rightarrow \ket{m_s = \pm1}$, generating an optically detected magnetic resonance (ODMR) spectrum at each camera pixel. Three permanent magnets, two static and one rotating, were used to apply a $\sim1$\,mT bias field to separate the resonances for the four different NV crystallographic orientations (Supplemental Material Sec. 2.2 \cite{supplement}). We measured the field along each NV axis from the Zeeman splitting of the corresponding resonance pair, yielding 2D magnetic field maps (Fig. \ref{fig:master_plot_all_particles}b).

\subsection{Sample mounting and temperature control}

The sample was cooled by two thermoelectric coolers (TECs) sandwiched between a thermally conductive sapphire slide and the sample mounting stage. The set temperature was monitored by a thermistor on the sapphire slide and stabilized via PID control. The \LCO particles were dropcast onto the NV-doped surface of the diamond sensor. The sensor was then mounted with the NV surface facing the sapphire slide and secured with thermal tape. We imaged through the 100\,\textmu m-thick diamond sample from below, allowing us to collect photoluminescence from the NV centers directly below the \LCO particles.

Changes in temperature cause a common-mode shift in each set of NV resonances (Supplemental Material Sec. 2.3.1~\cite{supplement}), allowing us to extract the diamond temperature from the ODMR spectra. All temperatures reported in this work are diamond temperatures. The \LCO particles discussed in this work appeared to be in contact with the diamond surface in brightfield images, but brightfield images cannot resolve a sub-micron gap, and control experiments show the diamond temperature is an imperfect proxy for the particle temperature (Supplemental Material Sec. 2.3.2~\cite{supplement}). Reducing the excitation power by one order of magnitude can shift the apparent N\'eel temperature higher by $\sim$3\,K, indicating a higher particle temperature during photoexcitation. The extent of the photothermal heating varies depending on the thermal contact between the \LCO and the diamond surface. 

\subsection{Data analysis}

Magnetic field maps were cropped to a region of interest around each particle, identified from low-temperature magnetic images. The bias field was subtracted from each map, leaving only the \textmu T-scale field from the particle. For the comparative magnetization curves, a baseline equal to the lowest $\ev{\abs{B_\mathrm{NV}}}$ in each dataset was subtracted from every point to account for variation in background level between datasets.

\section{Results and discussion}
 Magnetometry of eighteen particles revealed a range of behaviors: four showed no measurable stray field, nine showed magnetic signatures of similar magnitude above and below $T_N$, and five showed magnetic signatures appearing only below $T_N$, indicating a magnetic phase transition. Brightfield imaging confirmed the particles studied were isolated single crystallites rather than aggregates (Fig. S9 \cite{supplement}). The variety of behaviors observed likely reflects a combination of experimental and intrinsic factors. Particles showing no measurable stray field may have been in poor thermal contact with the diamond surface, preventing adequate cooling below $T_N$, or may simply have produced fields below our detection threshold; the peak fields we measure from the magnetically active particles are within an order of magnitude of our detection limit, estimated from the standard deviation of the noise in a signal-free image to be approximately 0.5\,\textmu T, so moderate particle-to-particle variation in moment, size or standoff from the diamond surface could place a particle below the threshold. 
 
 The magnetic signatures that are the same above and below $T_N$ likely arise from magnetic impurities or contamination within the sample, which would produce a constant background field over the experimental temperature range that is unrelated to the antiferromagnetic ordering transition. Given that the weak ferromagnetism in \LCO below the Néel temperature arises from slight spin canting in an antiferromagnetic state, the magnetization is several orders of magnitude smaller than typical ferromagnetic materials. Consistent with this expectation, the magnetic signals that do not change with temperature are typically stronger than the magnetic signals that undergo a clear transition and are often spatially localized to one region of a larger particle, suggesting that the signal originates from a small region within the \LCO microparticle (Supplemental Material Sec. 3 \cite{supplement}). These small ferromagnetic regions could be the result of a small amount of a CrO\textsubscript{2} impurity within the \LCO or Fe-bearing  contaminants introduced into the sample during synthesis or preparation for magnetic imaging \cite{garcia_magnetic_noise_sources}. Impurity phases are not observed in the ensemble XRD data (Fig. S7 \cite{supplement}), implying a very low concentration that is only visible using single-particle magnetic imaging. In the following, we focus our discussion on the particles that exhibit a magnetic phase transition near the expected Néel temperature.

\subsection{Observation of Néel transition in single-particle temperature sweeps}
\begin{figure*}[htb]
  \includegraphics[width=0.9\linewidth]{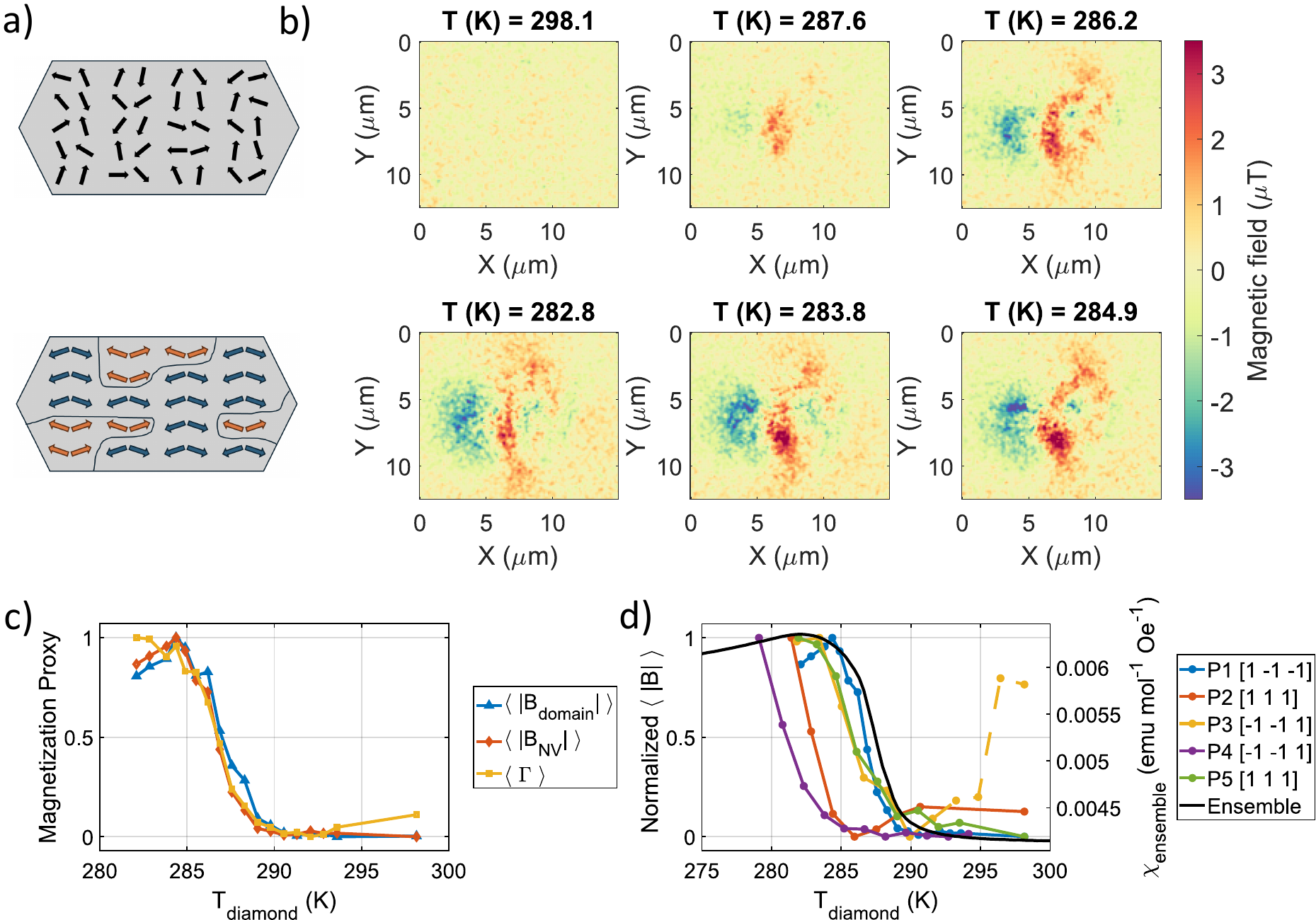}
  \vspace*{-0.1in} 
  \caption{(a) Cartoon of the spin orientations in a disordered paramagnetic state (top) and canted antiferromagnetic state (bottom). Blue and orange spins are canted downward and upward, respectively. (b) Representative magnetic field maps from a temperature sweep on Particle 1 collected using $\sim100$\,mW of laser power. Each map shows the projection of the magnetic field along the NV [1 -1 -1] axis directly beneath the particle as the temperature is swept down from room temperature. The $\sim$1\,mT bias field has been subtracted to only show the \textmu T-scale field from the \LCO particle. All temperatures have a $\pm0.4$\,K calibration uncertainty of the offset term in the linear function \cite{acosta_temperature_2010} that converts zero-field splitting of the NV center to temperature. (c) A comparison of the magnetization curves for the same Particle 1 temperature sweep shown in (b) derived using the three different methods described in the text: mean magnetic field over high-signal domains, $\ev{\abs{B_{\text{domain}}}}$; mean magnetic field over the whole particle, $\ev{\abs{B_{\text{NV}}}}$; and mean ODMR linewidth over the region of interest, $\ev{\Gamma}$. The $\ev{\abs{B_{\text{domain}}}}$ and $\ev{\abs{B_{\text{NV}}}}$ curves are taken from the projection along the NV [1 -1 -1] axis while the linewidth is extracted by fitting all NV axes to one $\Gamma$. We attribute the slight decrease of the magnetization at the low end of the temperature range to poor thermal contact between the particle and diamond, leading the \LCO temperature to plateau before the diamond temperature does. (d) Comparison of the normalized mean magnitude of $\ev{\abs{B_{\text{NV}}}}$ along the NV axis with the highest SNR, noted in the legend, for five different particles that displayed a clear Néel transition. The increase in signal for Particle 3 at high temperature is an artifact of low signal strength, rendering $\ev{\abs{B_{\text{NV}}}}$ susceptible to small changes in background noise level (raw magnetic field maps, unnormalized curves, and brightfield images are shown in the Supplemental Material Sec. 6 \cite{supplement}). The magnetic susceptibility measured via zero-field-cooled SQUID magnetometry at 100\,mT applied field on an ensemble sample of Yb:\LCO microparticles is shown in black (Supplemental Material Sec. 5).} 
 \label{fig:master_plot_all_particles}
\end{figure*}

Magnetic maps of the stray fields from five particles display a transition near the ensemble Néel temperature $T_N = 287$\,K. Selected magnetic maps of Particle 1 across a temperature sweep are shown in Fig. \ref{fig:master_plot_all_particles}b (full series in Supplemental Material Fig. S11 \cite{supplement}). In this experiment, the diamond temperature was swept from 298\,K to 282\,K. At room temperature, no stray fields are detectable in the magnetic field map. Near 288\,K, a small, spatially localized magnetic signal originating from the particle appears, signaling the onset of long-range magnetic order. This signal becomes progressively stronger as the temperature decreases until it plateaus or drops slightly at the lower end of the temperature range.

We identify three methods to characterize the onset of magnetic order from the single particle magnetic field maps. First, we directly measure the magnetic field vs. temperature in high-signal regions. To do this, we divide the magnetic image produced by a single particle into distinct spatial domains where the magnetic field is relatively uniform, then average the magnetic signal $B_{\text{domain}}$ from several domains to extract $\ev{\abs{B_{\text{domain}}}}$. These spatial ``domains'' do not necessarily precisely correspond to single magnetic domains due to the instrument's spatial resolution and to the 2D-projected nature of the measurement, in which distinct domains lying at different depths $z$ beneath the same region ($x, y$) are superimposed into a single signal. $B_{\text{domain}}$ is the most direct measurement of local variations in $M_{stag}$, but due to sample drift and low signal-to-noise in some domains, different regions underneath the same particle can produce inconsistent magnetization curves (Fig. S15 \cite{supplement}). 

Second, we quantify the magnetization by calculating the mean magnitude of the projection of the stray magnetic field along a given NV axis, $\ev{\abs{B_\text{NV}}}$, across the whole magnetic field map produced by a particle. This measurement averages over different magnetic features and is less affected by the choice of region of interest and therefore is robust against sample drift.

Third, the ODMR linewidth $\Gamma$ is used as a measure of magnetization, a technique which could also be used in cases where the magnetic field is varying rapidly in either space or time. For example, the ODMR linewidth increases when there is a magnetic field gradient across a single pixel because different NV centers within the pixel experience different Zeeman shifts. For particles with nano- or microscale magnetic features that are on the same length scale as the pixel size, an increased magnetization leads to a proportionate increase in the magnetic field gradient across a pixel. Due to this gradient, the ODMR linewidth increases in high-signal regions, and mean ODMR linewidth can be used as a proxy for average magnetic signal strength within a region of interest.

The relative magnetic transition curves for Particle 1 produced by all three methods
are shown in Fig. \ref{fig:master_plot_all_particles}c. The curves are in good agreement and all show the same sharp onset in magnetization near the Néel temperature. We use the mean field magnitude, $\langle |B_{\text{NV}}| \rangle$, dropping the subscript, as our standard proxy for particle magnetization when comparing across particles: averaging over every pixel in the region of interest reduces the contribution of per-pixel noise and sample drift, improving the SNR, and is a more direct proxy for magnetization than the ODMR linewidth, which reflects the field gradient rather than the field magnitude itself.

Comparative magnetization curves for five particles are shown in Fig. \ref{fig:master_plot_all_particles}d. All show a sudden onset of magnetization that increases sharply as the temperature decreases. However, the apparent Néel temperature varies significantly; Particles 1, 3 and 5 display Néel transitions within a couple degrees of the ensemble $T_N$, while Particle 2 and Particle 4 show no magnetic features until the diamond temperature is $\sim$\,$5$\,K colder. For these particles, the magnetization continues to rise over our whole cooling range, while the signals from Particles 1, 3 and 5 appear to reach an inflection point or plateau. 

Due to experimental limitations on \LCO particle temperature control and measurement, we cannot determine whether this spread reflects particle-to-particle heterogeneity in the Néel transition temperature or simply variations in experimental cooling efficiency. To test whether the measured diamond temperature accurately tracks the true \LCO temperature, we performed two diagnostic checks (Supplemental Material Sec. 2.3.2 \cite{supplement}). First, repeated imaging of a single particle over three hours showed no significant systematic drift in $\langle |B_{\text{NV}}| \rangle$, ruling out slow thermal equilibration as the source of the spread of the measured $T_N$. Second, reducing the excitation laser power, a known source of heating, by an order of magnitude increased the apparent $T_N$ by $\sim3$\,K, indicating that the particles dissipate heat less effectively than the diamond does, so the diamond's measured temperature systematically underestimates the true particle temperature. The apparent plateau in magnetization for some particles at the lowest temperatures likely has the same origin: over our measurement range ($0.97 < T/T_N < 1.03$), the staggered magnetization of a second-order transition should still be rising steeply, with no intrinsic saturation expected. Therefore, we suspect that the observed saturation in $\langle |B_{\text{NV}}| \rangle$ at the lower end of our temperature range reflects a limit on particle cooling rather than saturation of the order parameter.

The observed microtesla fields in the magnetic field maps are nearly three orders of magnitude smaller than the $\sim2.5$\,mT expected for a single domain, fully magnetized particle with a $0.5^\circ$ spin canting and magnetic moment normal to the particle face. The limited temperature range accessible in this experiment likely contributes to the difference between the measured and expected signal: the particle has not fully completed the Néel transition even at the coldest measured temperatures, and a thermal gradient across its poorly-thermalized surface could leave part of its volume too warm to have undergone the transition at all, so that the measured signal may originate from only a fraction of the particle's volume. Several additional factors may also contribute to this gap. First, geometry: the particle's easy axis is set by the local octahedral tilt and is generally not normal to the particle face, and a uniformly magnetized body with an in-plane moment produces stray field only near its edges. Second, we expect that there is a finite standoff between the particle and the diamond, since the particles do not have perfectly flat surfaces and were dropcast outside of a cleanroom environment, increasing the probability of dust particles causing a small standoff between the particle and surface. The sub-micron magnetic features we resolve indicate this standoff is itself sub-micron, since the standoff scales as roughly half the smallest resolvable feature size. However, even a small standoff could still cause significant attenuation: the field from a magnetization pattern of characteristic length $\lambda$ decays with standoff $z$ as $e^{-2\pi z/\lambda}$, so at $z\sim500$\,nm, this attenuates micron-scale features ($\lambda\sim1\,\mu\mathrm{m}$) by roughly $20\times$. Finally, the magnetic field measured in the diamond averages over multiple magnetic domains at varied z-heights that are projected into a single 2D plane at the diamond. Thermal cycling experiments suggest that these domains are micron-scale (Sec. \ref{sec:thermal_cycling}) so this cancellation will only be partial but will still contribute to attenuation of the magnetic signal. 

\begin{figure*}[htb]
    \centering
    \includegraphics[width=\linewidth]{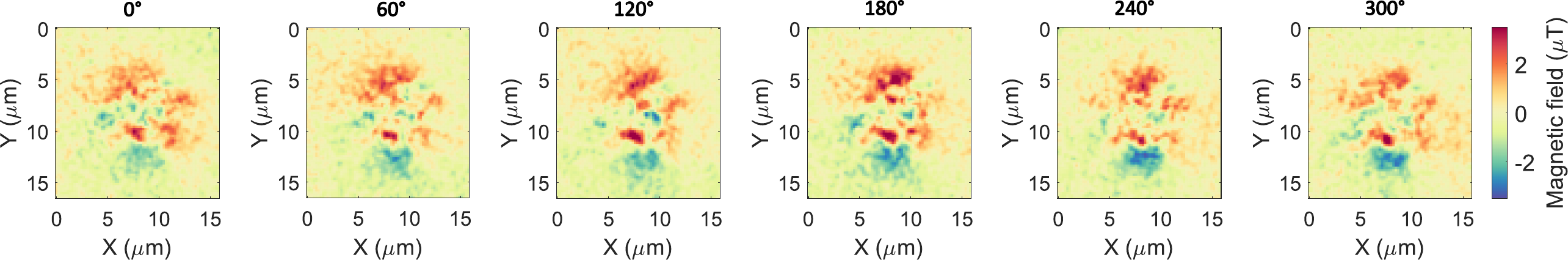}
    \caption{Particle 1 magnetic signal for varying angles (in degrees) of the $\sim1$\,mT in-plane component of the applied field while temperature is held constant to within 2 degrees of 284\,K.}
    \label{fig:anglesweep}
\end{figure*}

As further confirmation that the particles' behavior is consistent with weak ferromagnetism in a canted antiferromagnet, we investigated whether the domains remained aligned in the same orientation as the applied magnetic tweezer field was rotated. Using $K\sim10^4$\,J\,m$^{-3}$ and $M_\mathrm{net}\sim4\times10^3$\,A\,m$^{-1}$ (Sec.~\ref{sec:model}), the field needed to rotate the moment away from the easy axis, $B\sim K/M_\mathrm{net}$, is on the order of several tesla, orders of magnitude above the $\sim1$\,mT tweezer field used here, so a canted antiferromagnetic particle's moment should remain fixed while a paramagnetic particle's magnetization would rotate to align with the applied field. In this experiment, we held the temperature below $T_N$ at $284 \pm 2$\,K while we rotated the tweezer magnet (seen in Fig. \ref{fig:experimentalschematic}) over a $360^\circ$ range. The $B_\mathrm{NV}$ maps from the tweezer sweep are shown in Fig. \ref{fig:anglesweep}. We found that the stray magnetic field due to the \LCO magnetization remained fixed throughout the sweep, consistent with antiferromagnetic ordering.

\begin{figure}[htb]
  \includegraphics[width=\linewidth]{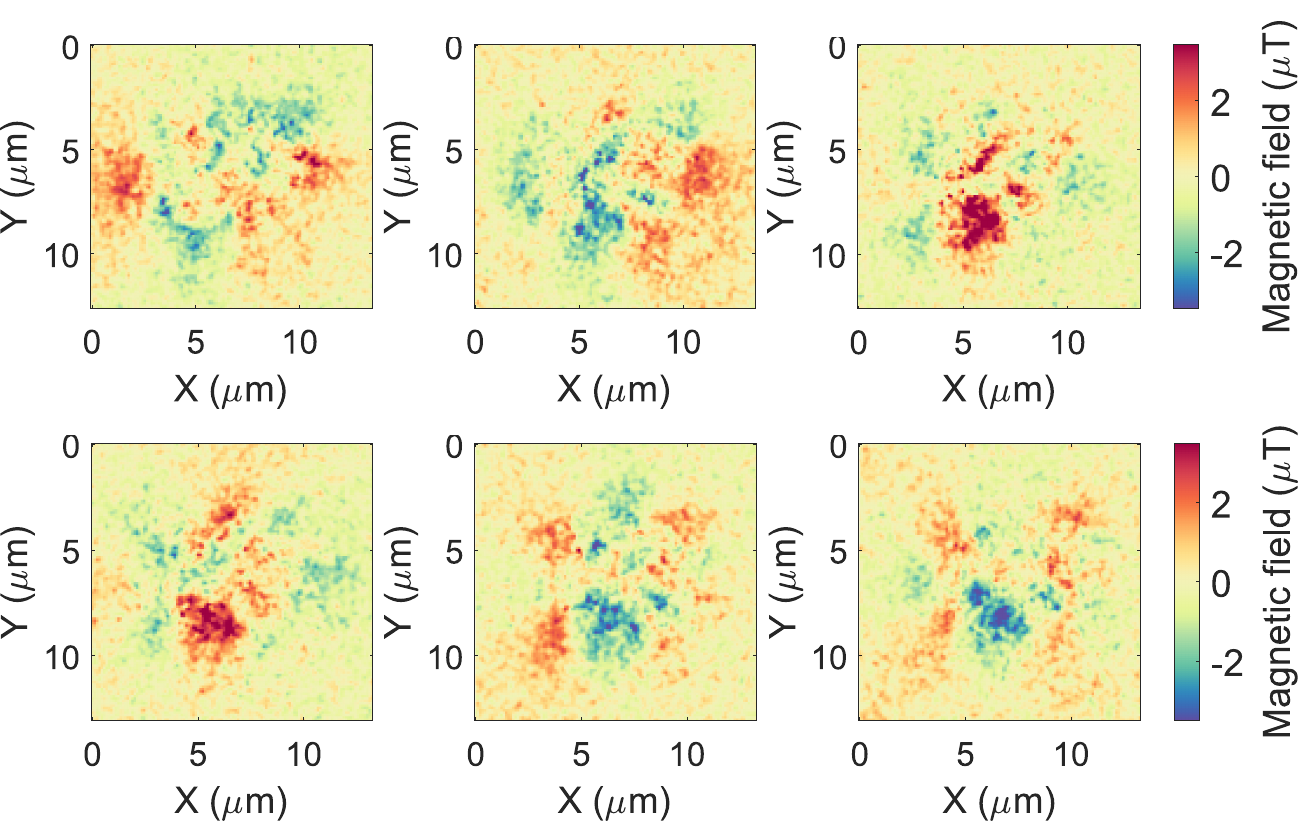}
  \vspace*{-0.1in} 
  \caption{Selected magnetic field maps from the temperature cycling experiment showing the projection of the magnetic field along the [1 -1 -1] NV axis underneath Particle 1. Each map was acquired at 283\,$\pm0.4$\,K, well below the onset of the Néel transition, and the temperature was brought up to 298\,K for 30 min between each acquisition to allow the particle to return to a disordered paramagnetic state. The magnetic field maps are all unique, highlighted by the variety of qualitative features in the three magnetic field maps shown in the top row, but similar features appear across multiple cooling cycles; for example, the three maps in the bottom row show a prominent triangular feature on the lower central region of the particle bordered by two weaker, oppositely magnetized lobes.}
 \label{fig:cycling}
\end{figure}

\subsection{Temperature cycling and magnetic domain formation }\label{sec:thermal_cycling}

The observation of apparent domain structure in Fig. \ref{fig:master_plot_all_particles}b prompts the question of whether we can identify multiple domains within our magnetic field maps and, if so, whether the domains are forming stochastically or deterministically. To experimentally address this question, we performed temperature cycling experiments: we measured the stray fields from a particle at low temperature, warmed it to 298\,K to reset it to a disordered state, then cooled it again. Cycling was repeated at four orientations of the in-plane applied field; we resolve no dependence of the resulting domain configuration on the field orientation (Supplemental Material Sec. 4~\cite{supplement}). Brightfield images confirmed the particle's position and orientation were unchanged across cycles, ruling out drift, within our instrument resolution, as the source of observed differences between repeated datasets on the same particle. 

A subset of the Particle 1 temperature cycling data are shown in Fig. \ref{fig:cycling} and the remaining data are available in the Supplemental Material Sec. 4 \cite{supplement} alongside temperature cycling data for additional particles in Fig.~S14. Two results emerge from this dataset. First, the field maps vary substantially from cycle-to-cycle, changing by more than an overall sign flip. This variation implies that each \LCO particle hosts multiple magnetic domains. Second, certain features recur across multiple cycles rather than a completely new configuration appearing every time. We estimated the magnitude of this recurrence via principal component analysis of the field maps, finding that five spatial patterns account for half of the cycle-to-cycle variation (Supplemental Material Sec.~4 \cite{supplement}). The reappearance of these features at consistent locations across thermal cycles suggests that the magnetic domains are bounded by fixed structural domain walls within the \LCO crystal, consistent with the absence of any change in the maps when the cooling rate is varied by an order of magnitude (Supplemental Material~\cite{supplement} Fig.~S13). The sign of each domain's magnetization, however, appears to be set independently at every cooldown. Because our field maps are two-dimensional projections of three-dimensional, multidomain particles, changes in the signs of neighboring domains can still alter the observed pattern in a given region, which accounts for the cycle-to-cycle variation superimposed on these recurring trends. Furthermore, the recurrence of a small number of distinct features suggests that the detected signal is dominated by domains on the micron scale: if the domains were significantly smaller than our imaging resolution, the magnetic field maps we measure would reconfigure seemingly stochastically at every cooldown. 

These results demonstrate that widefield NV magnetic imaging can resolve multidomain behavior within a single microcrystal and detect changes in the domain configuration. By repeatedly imaging the same particle, we can separate two effects that would otherwise be conflated in a measurement averaged over many particles or many cooldowns: the domain architecture, which is fixed and structurally pinned, and the domain magnetization, which is set stochastically at each cooldown.

\section{Conclusions and Outlook}
We have presented widefield NV-ODMR measurements of eighteen individual \LCO microcrystals at and near room temperature. Across these particles, we observed a range of behaviors: some particles showed no measurable magnetic signal, others produced a temperature-independent signal consistent with localized ferromagnetic impurities undetectable in ensemble XRD, and still others displayed a clear antiferromagnetic Néel transition. This split illustrates how single-particle measurements can reveal heterogeneity that bulk characterization averages away. Among the five transitioning particles, three key results emerged. First, stray magnetic fields materialized near $T_N$ and grew in magnitude upon further cooling, consistent with the onset of canted antiferromagnetic ordering. The particles transitioned at a range of temperatures and produced varying signal strengths. This variability is likely dominated by uneven thermal contact with the diamond sensor, but improved thermal control would let us test whether genuine differences in magnetic properties exist. Second, the magnetic signal remained fixed under a rotating $\sim$1\,mT applied field, consistent with canted antiferromagnetic ordering rather than paramagnetism. Third, repeated thermal cycling produced varied but partially recurring magnetic field patterns, consistent with a small number of structurally pinned antiferromagnetic domains that independently set their magnetization sign each cooldown.

This work demonstrates the utility of NV magnetic sensing for single-particle characterization of magnetic microcrystals: while ensemble magnetometry captures only the aggregate paramagnetic-to-antiferromagnetic transition in \LCO microcrystal samples, our single-particle approach can also detect small concentrations of impurities, measure potential particle-to-particle variability in transition behavior, and resolve the internal domain structure of individual particles, none of which bulk measurements can access. The number of particles characterized in this work was limited primarily by poor thermal contact rather than acquisition time, since data collection took only minutes per particle. Mounting the particles in a thermally conductive medium would improve thermal contact, while performing these measurements in a cryostat would additionally provide precise, wide-range temperature control and access to studying the Néel transition across a large set of orthochromites. Combining our existing magnetic imaging technique with improved thermal control would facilitate a survey of a large particle population and enable statistical studies of how morphology and synthesis conditions govern magnetic order and domain features in \LCO.

Beyond orthochromites, this work demonstrates the applicability of NV magnetometry to canted antiferromagnets more broadly, a class of magnetically rich materials that spans a wide range of material families. Widefield NV magnetometry can be applied across particles ranging from the nanoscale to hundreds of microns in size, allowing the same magnetic imaging technique to be used on particles in varied size regimes suited to specific applications, such as nanoscale thermometric probes, where resolving genuine particle-to-particle variation in $T_N$ would be essential. Furthermore, magnetic imaging on larger samples of particles would allow us to determine how variations in crystal structure, both within a sample and across different materials, affect antiferromagnetic domain structure and the Néel transition. This technique could also be extended to determine the types of domains within an antiferromagnet: because the ODMR spectrum yields the full vector field, for a particle with known morphology and standoff from the diamond sensor, the measured maps could be inverted to reconstruct the magnetization distribution itself rather than a single projection of the stray field~\cite{broadway_improved_2020}. Recovering the moment orientation within each domain would distinguish 180$^\circ$ domains, which reverse the canted moment along a common easy axis, from orientational domains belonging to distinct twin variants, and so link the antiferromagnetic domain pattern directly to the underlying crystallographic structure. Taken together, these results and future directions establish NV magnetometry as a powerful tool for single-particle characterization of structurally complex, magnetically interesting materials.

\section{Acknowledgments}
This research was supported by the NSF through the University of Washington Molecular Engineering Materials Center (MEM-C), a Materials Research Science and Engineering Center (MRSEC, DMR-2308979). Part of this work was conducted at the MEM-C Shared Facilities, also supported by NSF (DMR-2308979). 
 C.O.P. and R.T.S. gratefully acknowledge support through National Science Foundation Graduate Research Fellowships; this material is based upon work supported by the National Science Foundation Graduate Research Fellowship Program under Grant No. DGE-2140004. E.Q.W. was supported by an appointment to the Intelligence Community Postdoctoral Research Fellowship Program at The University of Washington administered by Oak Ridge Institute for Science and Education (ORISE) through an interagency agreement between the U.S. Department of Energy and the Office of the Director of National Intelligence (ODNI).

\bibliography{LaCrO3_pub}

\end{document}

\maketitle

\section{\LCO Synthesis Procedure}\label{SI_synthesis}

The salt mixture described in Sec.~III~A of the main text was ground using a mortar and pestle for 10 minutes before being transferred to a crucible and heated to $800\,^{\circ}\text{C}$ in a muffle furnace (MTI KSL-1100X-S-UL-LD) with a ramp rate of $6.5\,^{\circ}\text{C/min}$. The reaction was held at $800\,^{\circ}\text{C}$ for 3 hours before cooling to room temperature. The Yb:\LCO microcrystals were heavily aggregated and had to be scraped off the side of the crucible after being soaked for several minutes in 2~M hydrochloric acid. The reaction color at this stage was yellow due to the formation of potassium and sodium chromate. The microcrystals were then washed in water and centrifuged (5,000~rpm) three times, dissolving the sodium and potassium chromate and leaving a dark green product that could be suspended in ethanol.

It is important to note that this reaction produces potassium and sodium chromate(VI) complexes that are highly toxic, mutagenic, and carcinogenic. It is crucial to neutralize these to Cr(III) before disposal. This was done by adding an excess amount of iron(II) sulfate heptahydrate ($\geq$99.0\%, Sigma) to the waste acid and water from the reaction. The Fe(II) reacts with the yellow Cr(VI) solution and reduces it to a brown-red Cr(III) solution that can be disposed of in a waste container.

\section{Experimental Setup}
\subsection{Diamond sample} \label{SI_DiamondSample}
The diamond sensor consists of a 150 nm $^{15}\text{N}$-doped layer grown on an electronic-grade diamond substrate (Element Six) with source gas that had a $^{12}\text{C}$ isotopic purity of $>99.99$\%. Vacancies were formed by implantation with 25~keV He$^+$ at a dose of $5\times10^{11}$~ions/cm$^2$, followed by a vacuum anneal at 900$^\circ$C for 2~h and an O$_2$ anneal at 425$^\circ$C for 2~h for NV formation and charge state stabilization, respectively \cite{Kleinsasser2016}. The resulting NV ensemble has a density of $\sim\,1.7\times10^{16}$~cm$^{-3}$ and ensemble spin coherence time $T_2^* = 2.5~\mu$s \cite{Kazi2021, Kazi2024}.

\subsection{Widefield NV magnetometry scheme} \label{SI_experimental}
\begin{figure}[htb]
  \includegraphics[width=\linewidth]{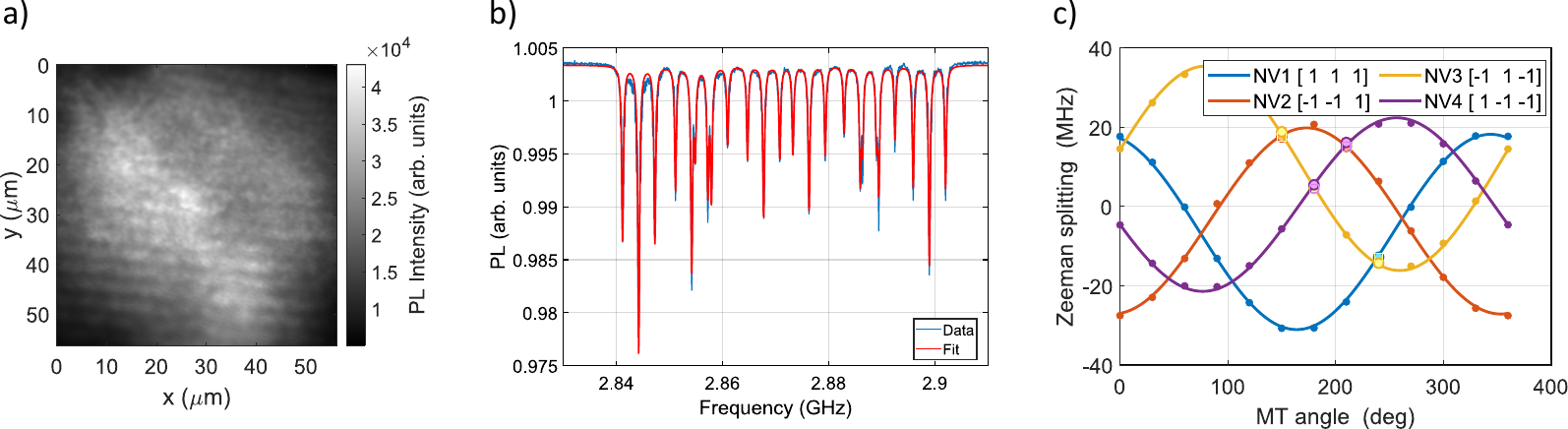}
  \caption{(a) Example NV photoluminescence image. (b) Example ODMR spectrum. (c) Calibration sweep of the four NV peak resonances vs. the angle of the applied in-plane field that is fit to a model of the applied field shape to identify which resonance corresponds to which NV crystallographic orientation at a given applied field angle.}\label{fig:pl_odmr_si}
\end{figure}
The NVs are excited with $\sim100$ mW of 532\,nm laser light focused to a $\sim25$\,\textmu m diameter spot using a 60\,x 0.95\,NA objective (Nikon MRD00605) and the photoluminescence is imaged onto an sCMOS camera (Hamamatsu OrcaFlash C13440-20CU). Each pixel corresponds to a 110\,nm by 110\,nm square on the sample, oversampling the $\sim400$\,nm optical resolution of the microscope. A Gaussian blur with a 2-pixel ($\sim220$\,nm) FWHM, smaller than the diffraction limit, is applied during image processing to reduce pixel-to-pixel noise. Optically detected magnetic resonance (ODMR) spectra are acquired using a frequency-swept, pulsed excitation sequence: a green laser pulse initializes the NV ensemble into the $\ket{m_s = 0}$ ground state, a resonant $\pi$ pulse transfers population to the $\ket{m_s = \pm 1}$ state, and a second green laser pulse reads out the resulting change in photoluminescence. The 4 ms camera exposure time is much longer than the $3.5~\mu \text{s}$ $\pi$-pulse and each pixel contains hundreds of NVs, so each image at a given microwave frequency is time averaged over many pulsed excitation cycles and spatially averaged over an NV ensemble.

The NV ensemble contains four crystallographic orientations, each contributing two spin transitions ($\ket{m_s = 0} \rightarrow \ket{m_s = \pm1}$). We simultaneously apply two microwave drive tones separated by the hyperfine splitting, $a_{hyp} = 3.05$\,MHz, yielding three resonance dips per electron spin transition: one at the unsplit resonance frequency $f_0$, one at $f_0 - a_{hyp}$, and one at $f_0 + a_{hyp}$. Transitions for the four NV crystallographic orientations are split by the applied $\sim1$\,mT bias magnetic field. Sweeping the microwave frequency across the full NV ODMR spectrum therefore yields a total of 24 resonances (Fig. \ref{fig:pl_odmr_si}b). The spectrum is fit to a sum of 24 Lorentzians, producing 8 center frequencies corresponding to the two spin transitions for each of the four NV orientations. In the low magnetic field limit, each NV orientation experiences a frequency shift given by the projection of the magnetic field along the NV axis. We then convert the Zeeman shift to a magnetic field projection by taking the difference between each set of $\ket{m_s = \pm1}$ resonances, $f_+ - f_- = 2\gamma_\text{NV} B_\parallel$, where $f_+$ and $f_-$ refer to the resonant frequencies extracted from the ODMR spectrum of the $\ket{m_s = 0} \rightarrow \ket{m_s = +1}$ and $\ket{m_s = 0} \rightarrow \ket{m_s = -1}$ transitions, respectively, and $\gamma_\text{NV} = 28~\text{MHz}/\text{mT}$ is the NV gyromagnetic ratio. The NV orientation that corresponds to each resonance is identified via a calibration scan in which we sweep the in-plane applied field direction and fit the resulting angle-dependent Zeeman shifts to extract the alignment between the NV frame and the external magnetic field (Fig. \ref{fig:pl_odmr_si}c). 

\subsection{Temperature measurements}

\subsubsection{Measurement of the diamond temperature}
\label{SI_TemperatureMeasurement}
Temperature is extracted by fitting the NV zero-field splitting (ZFS), $f_\text{ZFS} = (f_+ + f_-)/2$. The ZFS is shifted by crystal strain as well as temperature, which varies between different NV orientations and spatially across the sample. Therefore, for each experiment we calculate the relative temperature by measuring the change in ZFS relative to that at room temperature ($\text{T} = 298.15$ K) for each NV orientation and then averaging across all measured NV orientations. The ZFS shift is then converted to a temperature using the value determined by Acosta et al., -$74.2(7)~\text{kHz}/\text{K}$ \cite{acosta_temperature_2010}. Given the small temperature range probed in this experiment ($\sim 15$ K), we expect the dominant contribution to the temperature uncertainty to be the calibration error from determining the $298$ K ZFS. We estimate this uncertainty from the variation in the ZFS over fourteen datasets collected at $\text{T} = 298.15$ K, added in quadrature across four NV orientations, yielding a calibration uncertainty of $\pm$0.4 K.

\subsubsection{Laser heating and thermal contact between microparticles and diamond}
\label{SI_LaserHeating}
\begin{figure}[htb]
  \includegraphics[width=\linewidth]{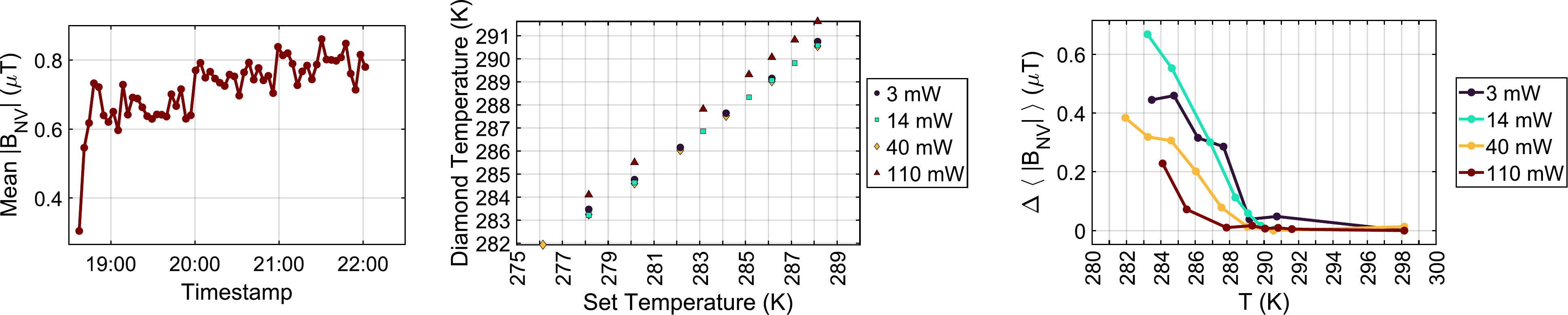}
  \caption{Left: Particle 1 $\ev{\abs{B_{\text{NV}}}}$ vs. timestamp over three hours while the diamond temperature was held at $283.4\pm0.4$\,K. Excepting the small jump at 20:00 which we attribute to particle movement, $\ev{\abs{B_{\text{NV}}}}$ remains approximately constant, implying that there is no systematic drift in microparticle temperature over time. Center: Diamond temperature vs. set temperature during temperature sweeps on Particle 1 for varying laser powers. Right: $\ev{\abs{B_{\text{NV}}}}$ curves extracted from the same dataset shown in the center plot. Note that this data was collected a month after the Particle 1 temperature sweeps discussed in the main text, and the particle appeared in brightfield images to have shifted in the intervening time. Therefore, these $\ev{\abs{B_{\text{NV}}}}$ curves do not match with the other Particle 1 data, likely due to a systematic shift in the thermal contact between the \LCO particle and the diamond sample.}\label{fig:laser_heating}
\end{figure}

The efficacy of the \LCO microparticle cooling in our experiment is limited primarily by the thermal contact between the uneven particle surface and the diamond. This problem is exacerbated by the continual addition of heat to the diamond and particles from the green laser used to excite the NVs. To investigate the effects of these factors on our data, we performed two types of control experiments (Fig. \ref{fig:laser_heating}). 

In the first experiment, we took repeated scans of the same particle over the course of three hours, the same time scale as the slowest experiments discussed in the main text, and tracked $\ev{\abs{B_{\text{NV}}}}$ vs. time (Fig. \ref{fig:laser_heating}). With the exception of a jump in $\ev{\abs{B_{\text{NV}}}}$ around one hour into the experiment (20:00), which we attribute to a small shift in the particle orientation during autofocusing, $\ev{\abs{B_{\text{NV}}}}$ remained approximately constant after the initial cooldown period, with a potential small linear drift with a magnitude smaller than the random fluctuations. This implies that over the course of our experimental timescale, the particle is in a thermal near-steady state and that there is no significant systematic temperature drift over the course of our experiments.

In the second experiment, we studied the impact of optical power on the measured N\'eel temperature. Because N\'eel temperature is determined from $\langle|B|\rangle$ as a function of diamond temperature, a laser-power-dependent shift in the diamond's own temperature would not, on its own, produce any change in the observed N\'eel temperature: if the particle tracked the diamond's temperature change exactly, the transition would simply occur at the same point on the diamond-temperature axis. However, we observed that decreasing the laser power from $110$\,mW to $40$\,mW decreased the diamond temperature by approximately 1\,K, yet increased the observed N\'eel temperature by nearly 2\,K, indicating that the particle and diamond temperatures do not shift together. Further decreasing the optical power to $14$\,mW caused an additional $\sim1$\,K shift in N\'eel temperature with no measurable accompanying change in diamond temperature, again indicating a particle-specific response. These results support our supposition that the diamond temperature is an imperfect proxy for \LCO microparticle temperature and imply that the microparticles are more affected by laser heating than the diamond is. We attribute this difference to a combination of poor thermal contact between the particles and the diamond, limiting their ability to dissipate the heat added by the laser, and differences in laser absorption between the diamond and the \LCO. 

\section{Magnetic Impurities}\label{SI_magneticimpurities}
\begin{figure}[htb]
  \includegraphics[trim=5cm 3cm 3cm 3cm, clip=true, width=\linewidth]{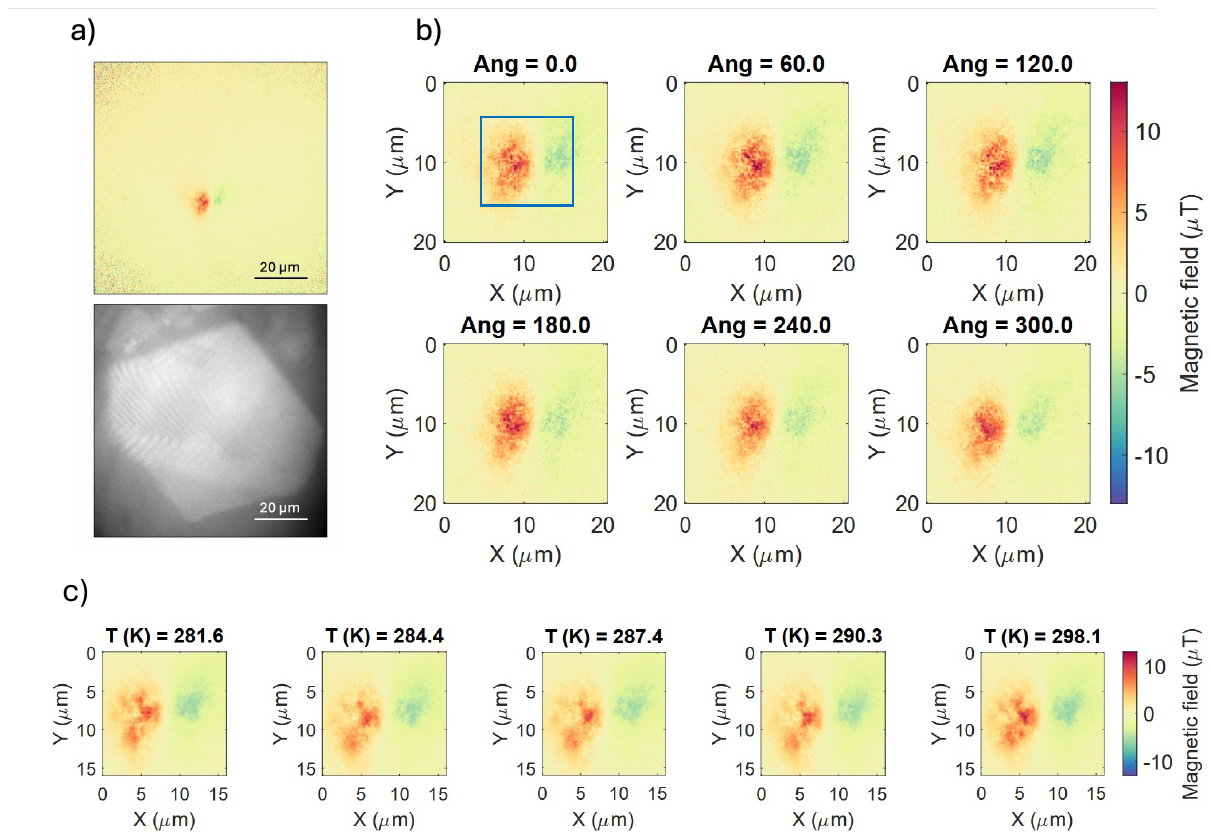}
  \caption{(a) Room temperature magnetic field map of a particle with a magnetic signature not exhibiting a phase transition, and brightfield image of the particle. The magnetic field is the projection along the NV [1 -1 -1] axis. (b) Magnetic signature over varying applied field angles. The crop region used to determine $\langle |B_{\text{NV}}|\rangle$ in Sec. \ref{SI_magneticimpurities} of the Supplemental Material is outlined in blue. The region in (b) outside of the blue box is used to determine the noise floor for the magnetic field map for background subtraction. (c) Magnetic field map of the particle at different temperatures. The magnetic signature is nearly invariant over temperatures above and below the \LCO Néel temperature.}\label{fig:no_transition}
\end{figure}

In particles that show a temperature-independent magnetic signature, the magnetic signal is typically a two- or four-lobed pattern that resembles the field of a point dipole and is often localized to a small area of the particle. A representative example is shown in  Fig. \ref{fig:no_transition}, which depicts a small two-lobed dipole pattern under a much larger particle; the dipole-like field does not change significantly over a range of temperatures. In contrast, the particles that display a Néel transition typically produce more complicated magnetic field patterns that appear to originate from many different parts of the particle. 

We confirmed that the temperature-independent signals are consistent with ferromagnetic ordering by sweeping the angle of the $\sim1$\,mT applied magnetic tweezer field. The magnetic signal remained approximately constant as the applied magnetic field angle was varied over 360$^\circ$ (Fig. \ref{fig:no_transition}b). This suggests that the temperature-independent signals may originate from localized ferromagnetic impurities in the \LCO microcrystals with a higher Curie temperature than the \LCO Néel temperature. One candidate for the impurities is  CrO\textsubscript{2} impurities formed during the molten salt synthesis.

Ferromagnetic impurities or inclusions are expected to have a net magnetization orders of magnitude larger than that of a canted antiferromagnet. As a coarse check of whether our data are consistent with this expectation, we compared the mean $\ev{\abs{B_{\text{NV}}}}$ of the five particles with a magnetic phase transition to that of the five non-transitioning particles for which we collected full temperature sweeps. We removed the background contribution from our noise floor by subtracting the mean $\ev{\abs{B_{\text{NV}}}}$ computed outside the cropped region of interest (ROI) around each particle from the mean $\ev{\abs{B_{\text{NV}}}}$ computed within the ROI. The mean field magnitude of the five temperature-independent signatures at room temperature is $2.9\pm1.4$\,\textmu T, roughly six times stronger than the mean low-temperature field magnitude of the five particles displaying a Néel transition, $0.5\pm0.5$\,\textmu T. This difference is consistent with small, localized ferromagnetic impurities, which may lie anywhere within a particle and thus considerably farther from the diamond surface than the material undergoing the Néel transition, which must be in close thermal contact with the diamond to cool effectively.

\clearpage

\section{Temperature Cycling and Domain Feature Analysis}
\label{SI_DomainAnalysis}

\begin{figure}[htb]
  \centering
  \includegraphics[trim=1cm 6.5cm 0.5cm 4.6cm, clip=true, width=\linewidth]{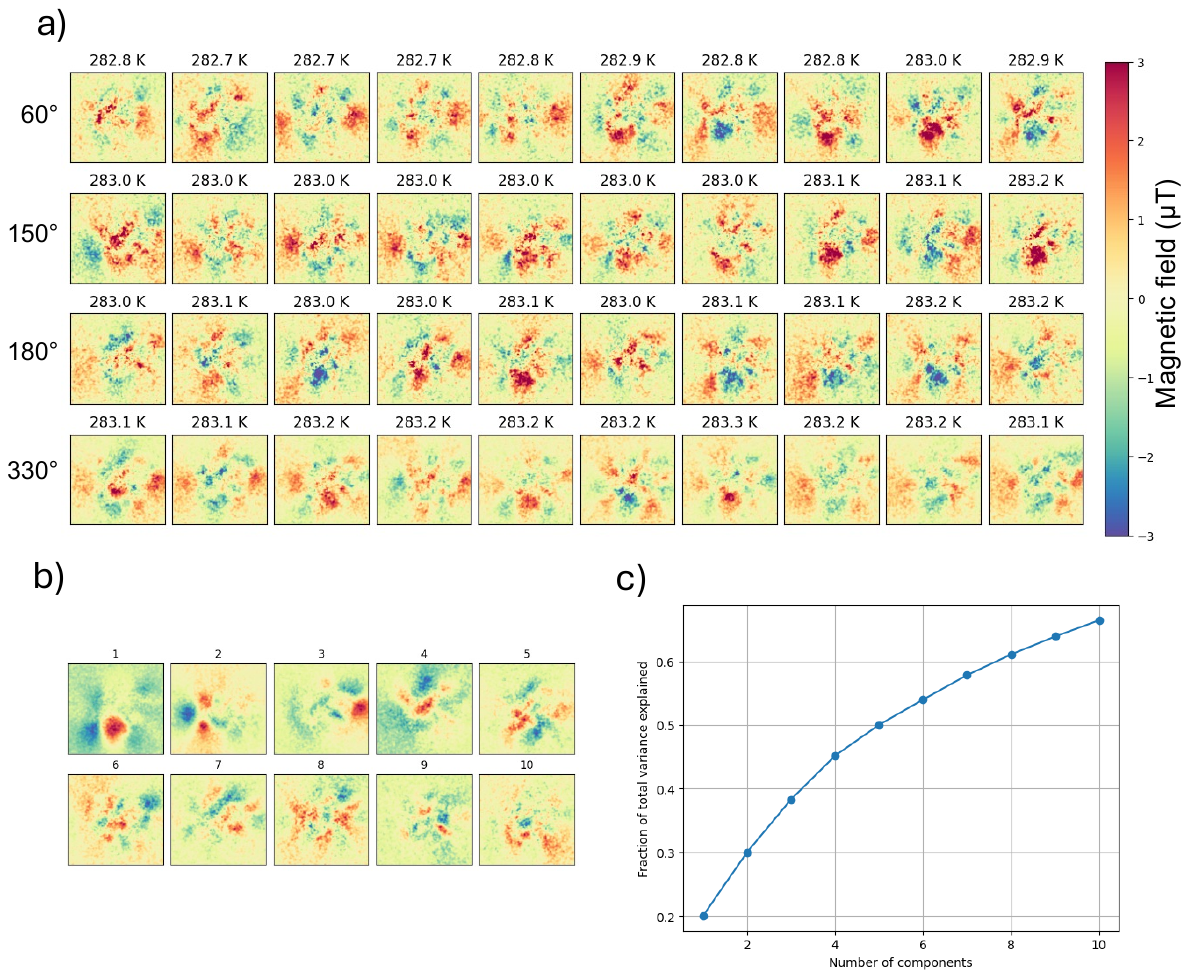}
  \caption{(a) Temperature cycling data on Particle 1 at four applied field angles, with the projection of the stray field along the [1 -1 -1] NV axis plotted. The images are $11.6 \times 12.1$ \textmu m. (b) First 10 components obtained from PCA. (c) Fraction of the total variance explained vs. the number of components used to reconstruct the images.}\label{fig:cycling_data_supp_p1}
\end{figure}

\begin{figure}[htb]
  \centering
  \includegraphics[trim=2cm 3cm 2cm 3cm, clip=true, width=\linewidth]{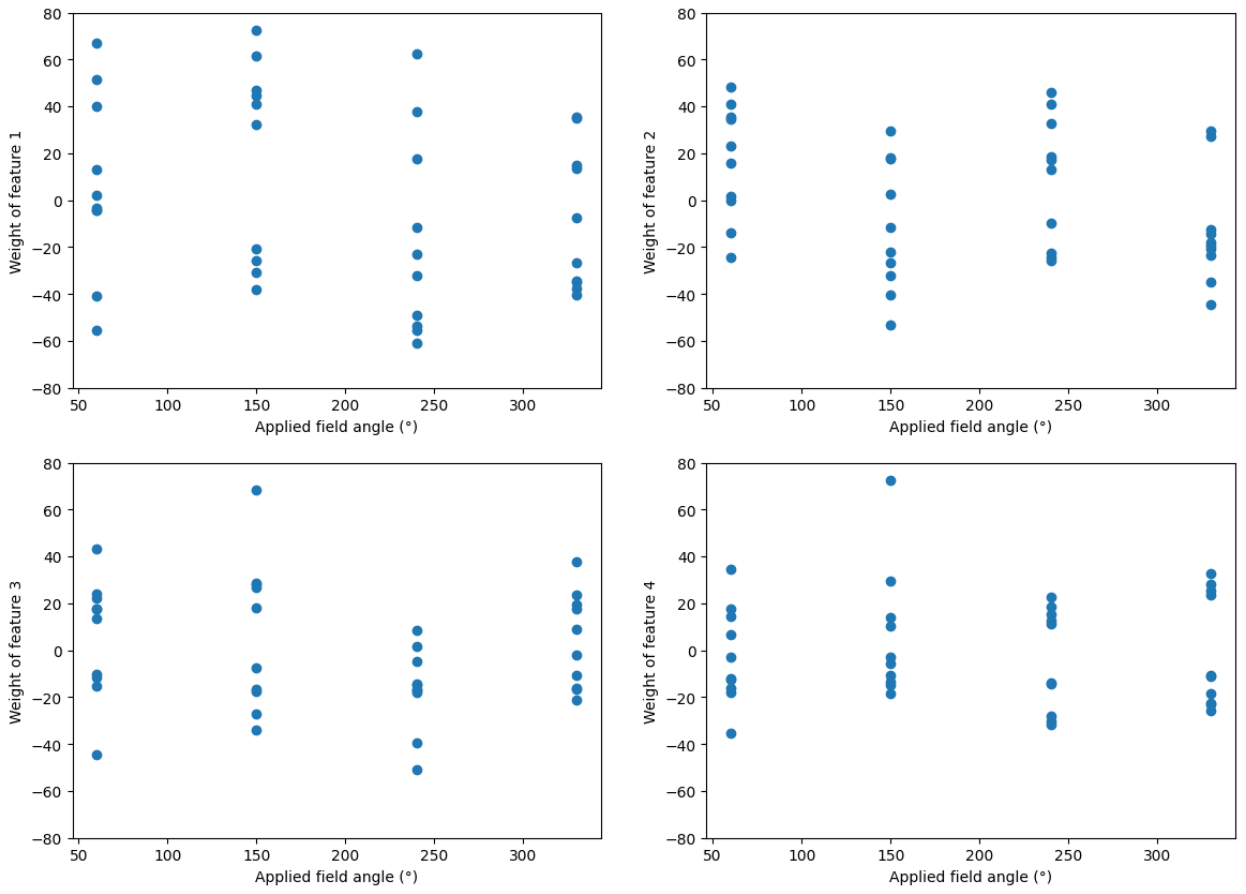}
  \caption{Projections of Fig. \ref{fig:cycling_data_supp_p1} cycling data onto the first four principal components obtained by PCA, separated by applied field angle.}\label{fig:principal_components_analysis_supp}
\end{figure}

Temperature cycling data was taken on Particle 1 at applied field angles 60$^\circ$, 150$^\circ$, 180$^\circ$, and 330$^\circ$, where ten cycles were carried out for each angle (Fig. \ref{fig:cycling_data_supp_p1}a). It is apparent that, while the qualitative pattern generally changes between cycles, certain structural features recur at consistent locations across cycles.

As a compact method of summarizing the spatial features that vary most across the dataset, we performed principal component analysis (PCA) on the 40 magnetic field images. Each $105\times110$ pixel image was flattened into an $11550$-dimensional vector, and the resulting vectors were arranged into a $40\times11550$ data matrix $X$, with one row per image. After subtracting the mean of each column (i.e. the per-pixel mean across the 40 images) from $X$, we computed the PCA via scikit-learn's implementation, which decomposes the centered data matrix by singular value decomposition, $X = SV^\top$, where the columns of $V$ form an orthonormal basis of spatial patterns ordered by the fraction of the dataset's total variance each one explains, and $S$ gives each image's projection onto every such pattern. The first ten of these patterns, reshaped back into $105\times110$ images, are shown in Fig.~\ref{fig:cycling_data_supp_p1}b. The first five components account for half of the total variance in the data (Fig.~\ref{fig:cycling_data_supp_p1}c), indicating that a small number of spatial patterns describes much of the cycle-to-cycle variation.

In addition, the projections of each image onto the four most significant features are plotted against the applied field angle to determine whether the applied field angle has an effect on the magnetic domain configuration (Fig. \ref{fig:principal_components_analysis_supp}). The distributions of the projections are comparable across the four applied field angles, with the scatter within a given angle similar in magnitude to the variation between angles. We therefore do not resolve a systematic dependence of the domain configuration on the applied field angle. With ten cycles per angle these data cannot exclude a weak bias, but they are consistent with the sign of each domain's magnetization being set independently at each cooldown.

\section{Ensemble Measurements}

Here we present ensemble measurements of \LCO samples from the same synthesis as those measured with NV magnetometry in the main text.

\label{SI_Ensemble}
\begin{figure}[htb]\centering
  \includegraphics[width=\linewidth]{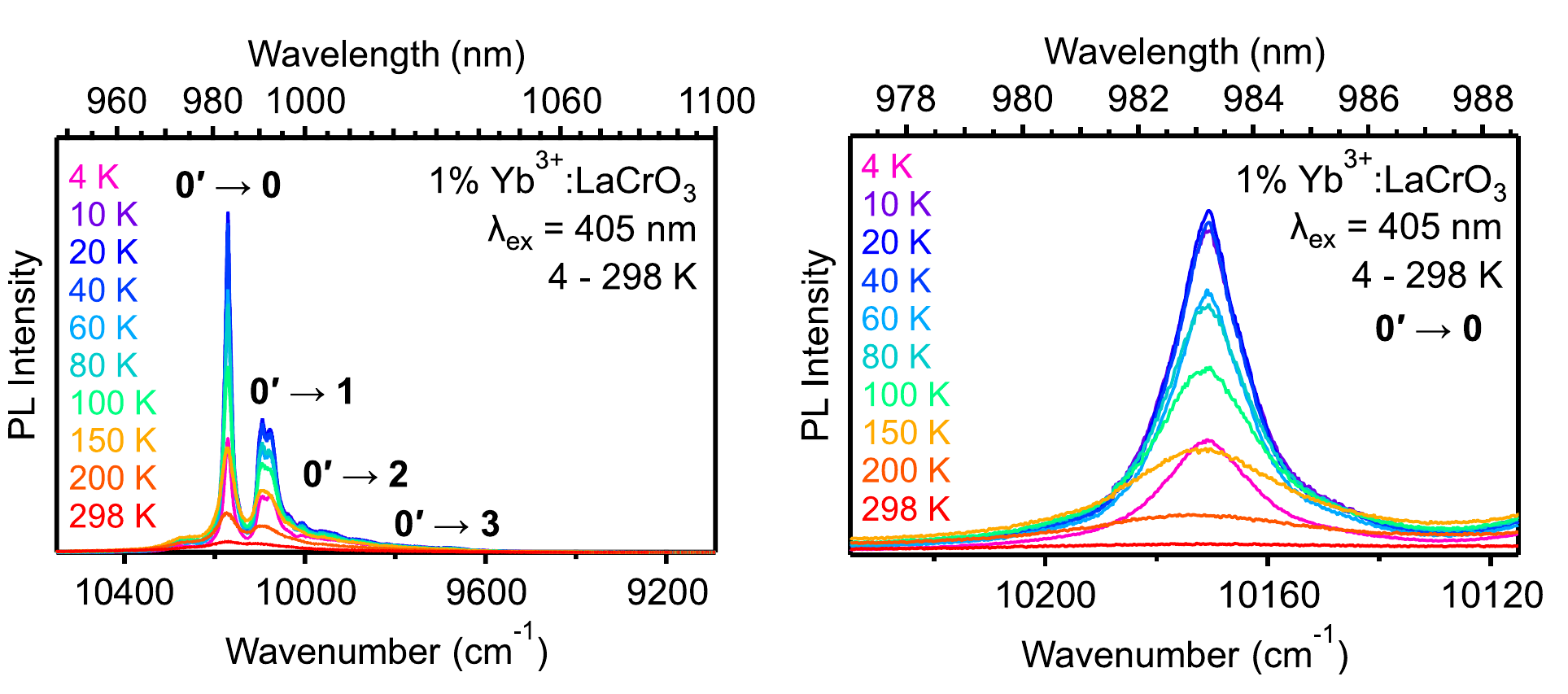}
  \caption{Ensemble photoluminescence spectra of the 1\% ytterbium that is doped into the \LCO. Peaks are labeled $0' \rightarrow n$, denoting emission from the lowest crystal-field sublevel of the Yb$^{3+}$ $^2F_{5/2}$ excited state to each crystal-field sublevel of the $^2F_{7/2}$ ground state ($n = 0$--$3$). We expect to see a characteristic exchange splitting signal in the $\text{Yb}^{3+}$ spectrum from one peak into two arise as the material passes through the Néel transition \cite{tzanetopoulos_luminescent_2026}. However, we are unable to resolve this splitting even at 4\, K, which we attribute to the Yb$^{3+}$ lines being too broad to resolve the exchange splitting. This broadening may result from sample-specific defects and surface/core $\text{Yb}^{3+}$ heterogeneity.}\label{fig:yb_pl}
\end{figure}

\begin{figure}[htb]\centering
  \includegraphics[width=0.5\linewidth]{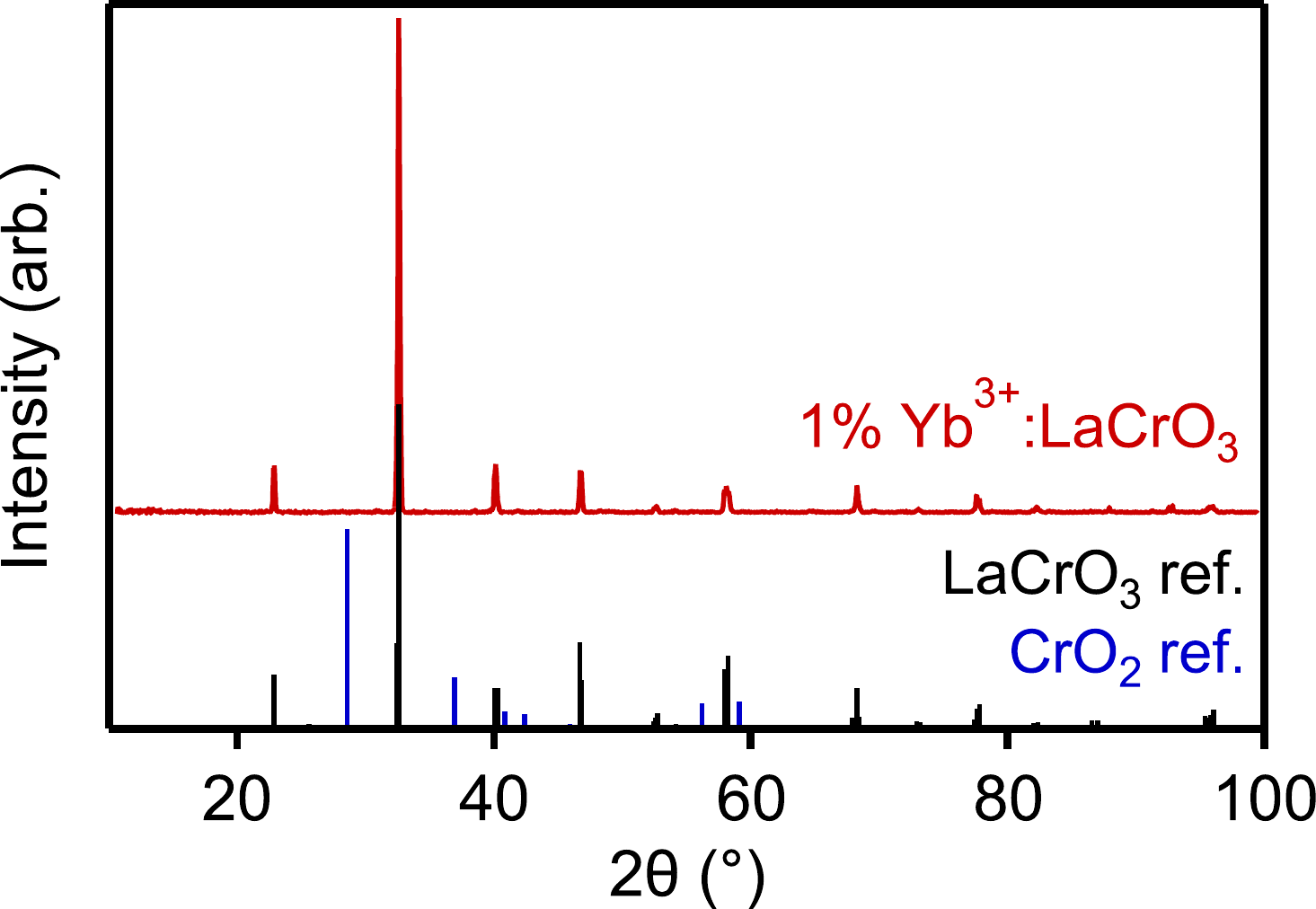}
  \caption{X-ray diffraction (XRD, Bruker D8 Discover) spectrum of an ensemble of the 1\% Yb \LCO microcrystals using Cu K-$\alpha$ radiation ($\lambda = \SI{1.5406}{\angstrom}$). The microcrystals were dropcast onto a silicon wafer for XRD analysis. The spectrum is consistent with the \LCO reference. For comparison, we include the CrO\textsubscript{2} reference which is one potential ferromagnetic contaminant that could be formed during the \LCO synthesis. No peaks match, indicating that if CrO\textsubscript{2} contamination is present, it is at a concentration below the detection limit of this XRD measurement.}\label{fig:xrd_data}
\end{figure}

\begin{figure}[htb]\centering
  \includegraphics[width=\linewidth]{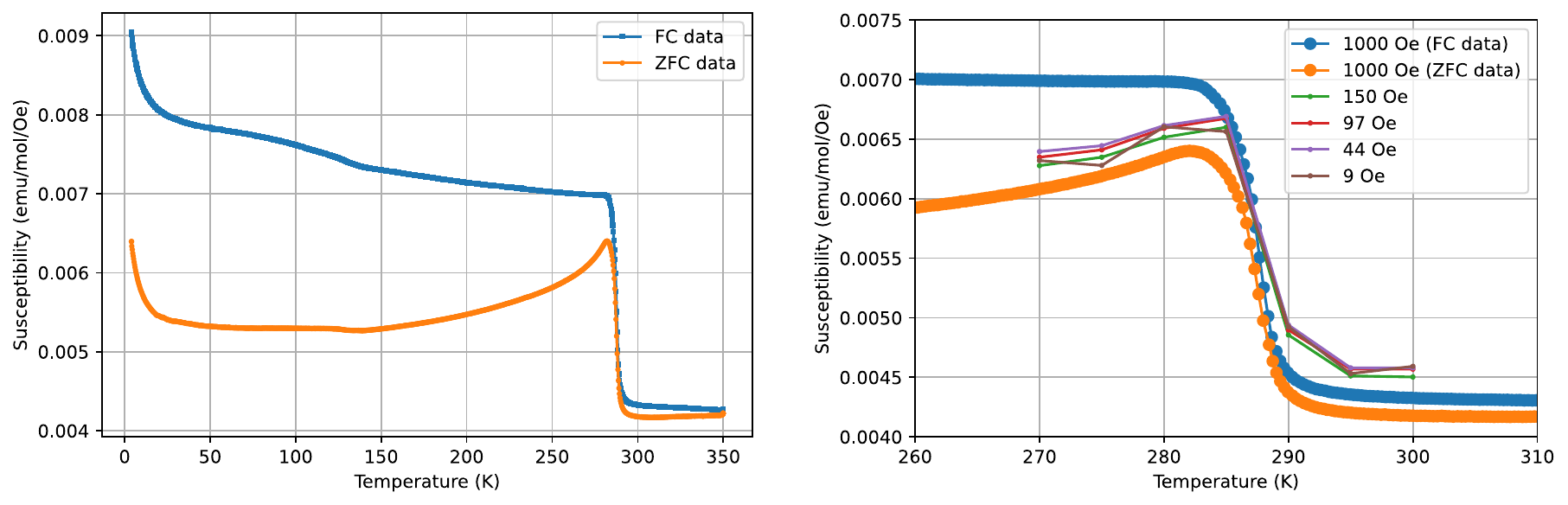}
  \caption{Field-normalized SQUID susceptibility measurement of field-cooled (FC) and zero-field cooled (ZFC) on a 3.6\,mg sample of Yb:\LCO. On the right, SQUID measurements of partial temperature sweeps near the Néel temperature are shown for varying applied fields, confirming that the susceptibility remains approximately constant at applied fields down to the field used in our single particle experiments (9\,Oe = 0.9\,mT; the applied field in the single particle experiments is $\sim$1\,mT). The measurements were performed on a Yb:\LCO microcrystalline powder that was mixed with a small amount of rubber cement and adhered to a quartz sample holder using a Quantum Design MPMS$^\text{\textregistered}$3 SQUID magnetometer (MEM-C Shared Facilities) operating in the vibrating sample magnetometer mode. }\label{fig:SQUID_FC_ZFC_Full_Temp}
\end{figure}

\FloatBarrier
\section{Néel transition particles}
\label{SI_NeelParticles}

\begin{figure}[htb]
  \includegraphics[width=\linewidth]{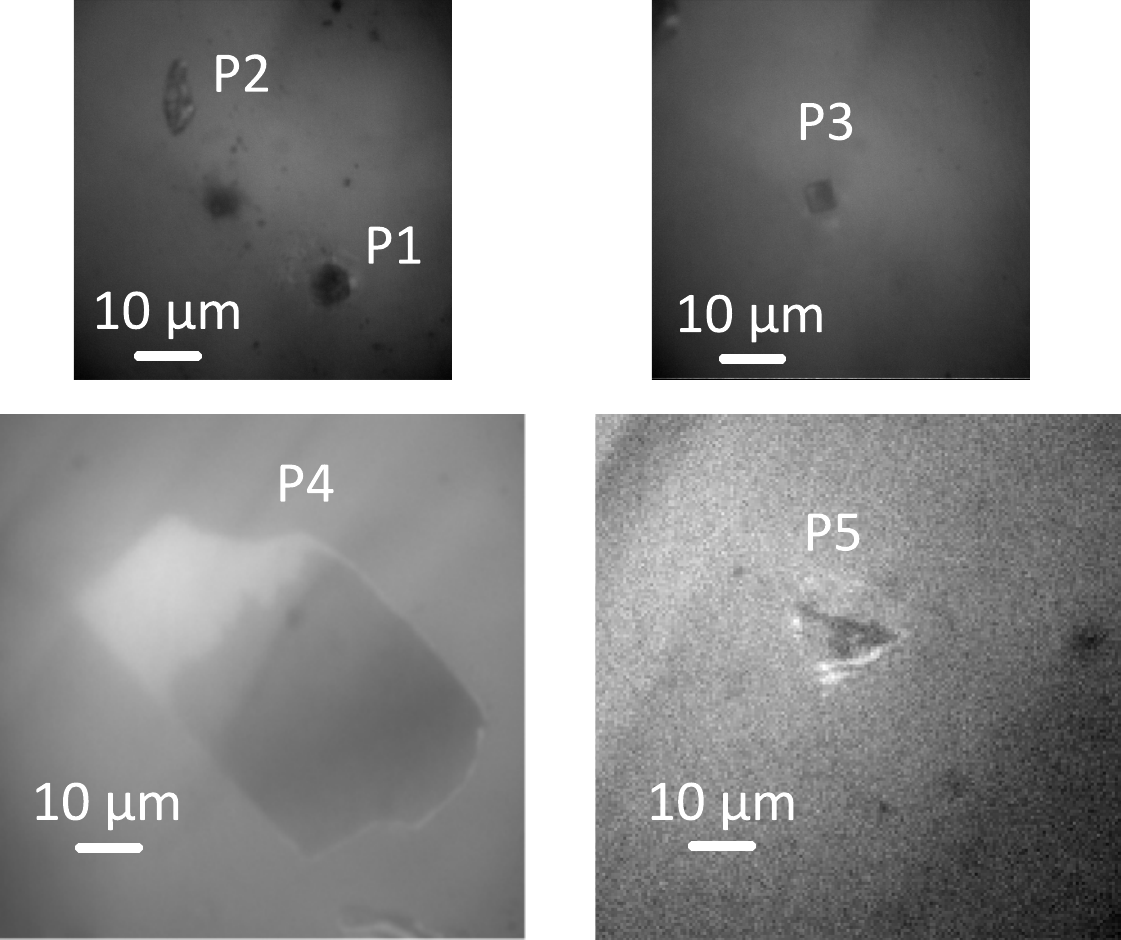}
  \caption{Brightfield images of Particles 1 through 5.
  }\label{fig:brightfield_composite}
\end{figure}

\clearpage

\FloatBarrier

\subsection{Magnetic imaging experiments}

\begin{figure}[htb]\centering
  \includegraphics[width=\linewidth]{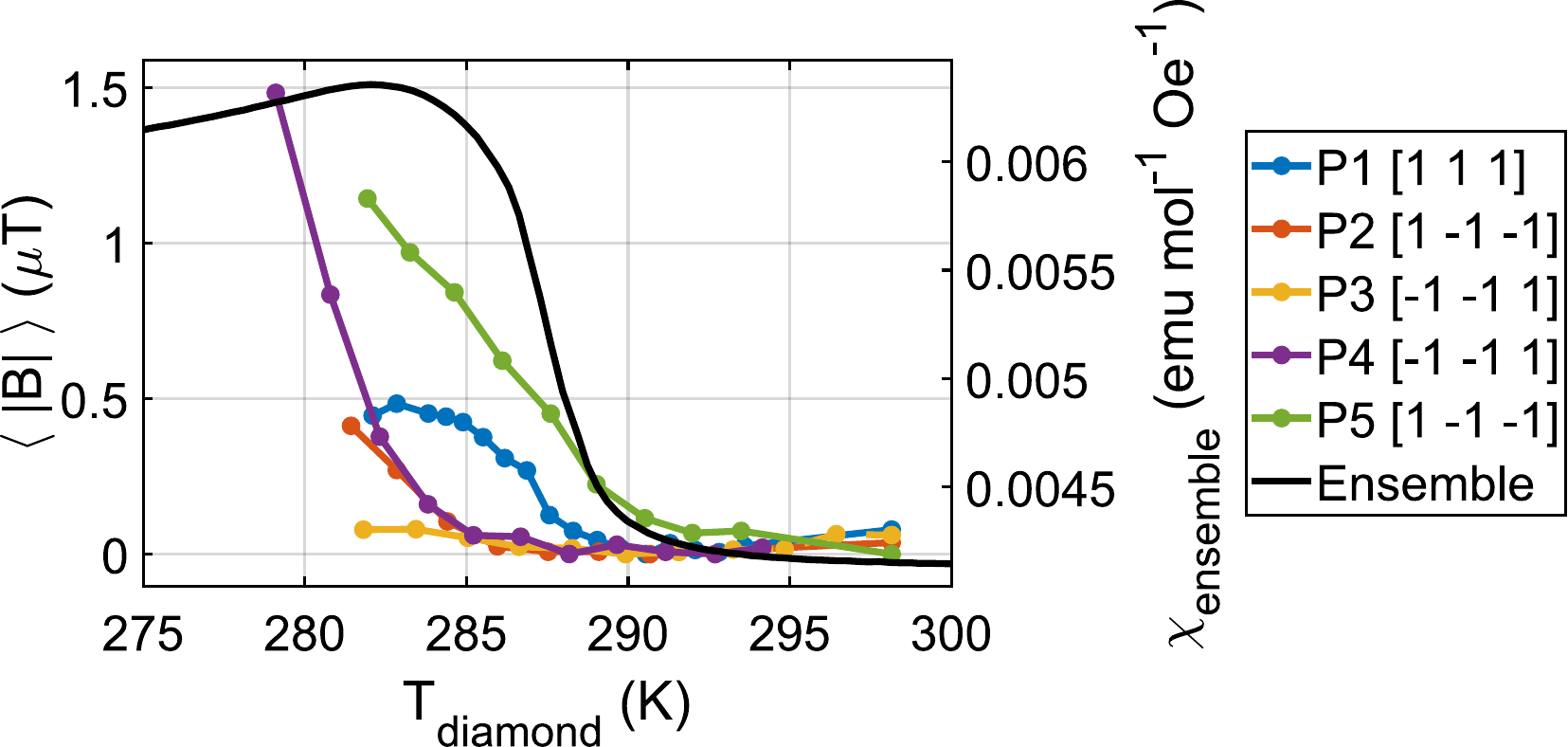}
  \caption{Non-normalized $\ev{\abs{B_{NV}}}$ curves for the particles that display a N\'eel transition. The data are baseline subtracted by subtracting the lowest $\ev{\abs{B}}$ value across the dataset from every datapoint to account for variations in the background noise level between datasets. For each particle, the data for the projection along the NV orientation that shows the greatest change in magnetization, noted in the legend, is plotted. The ZFC SQUID susceptibility data taken at 100\,mT is also included.}\label{fig:magnetization_not_normalized}
\end{figure}

\begin{figure}[htb]
  \includegraphics[width=\linewidth]{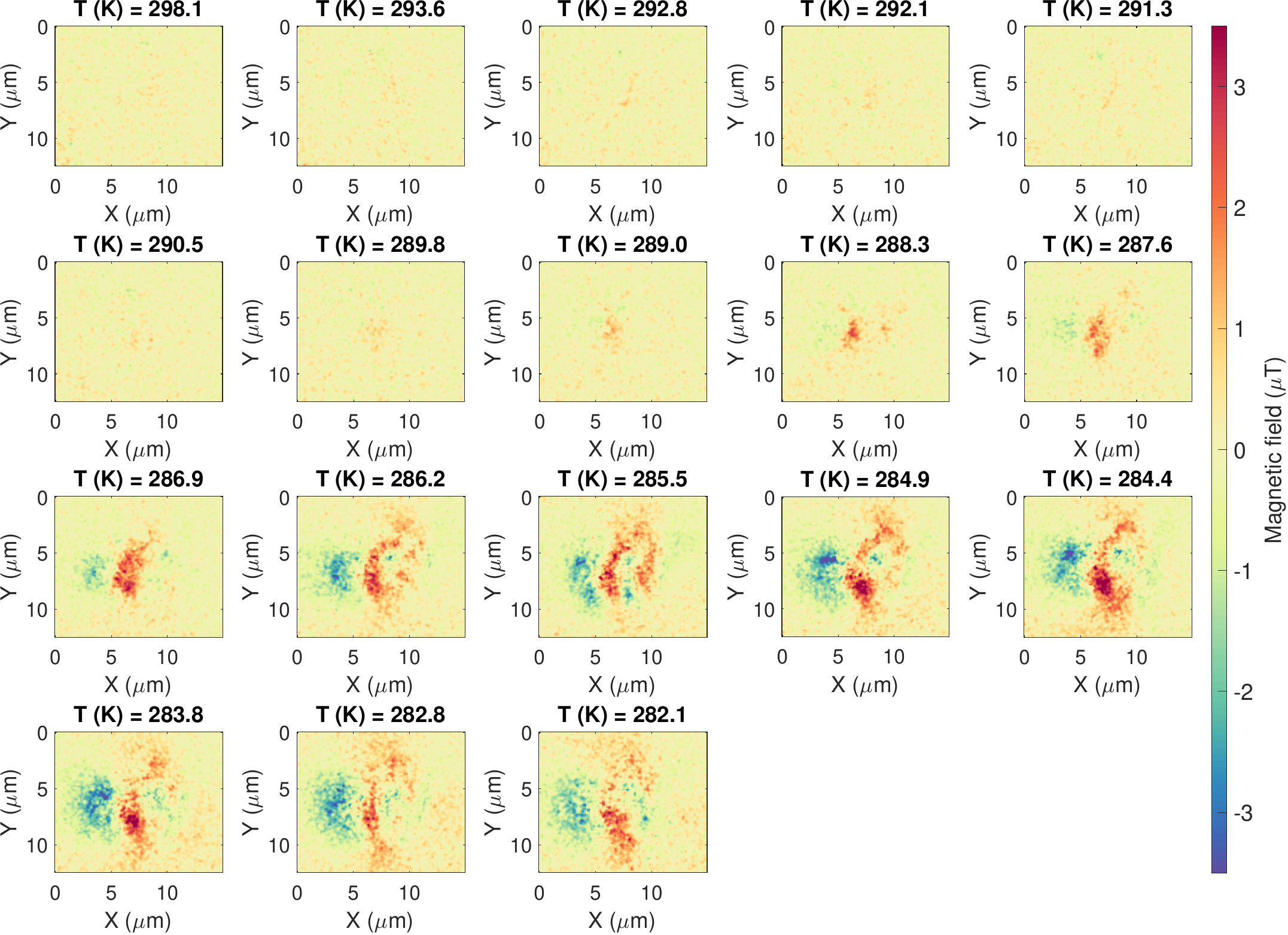}
  \caption{Full magnetic field map data, projected along the NV [1 -1 -1] axis, for the Particle 1 temperature sweep used to compute the magnetization curves presented in the main text.}\label{fig:particle1_extended_sweep}
\end{figure}

\begin{figure}[htb]
  \includegraphics[width=\linewidth]{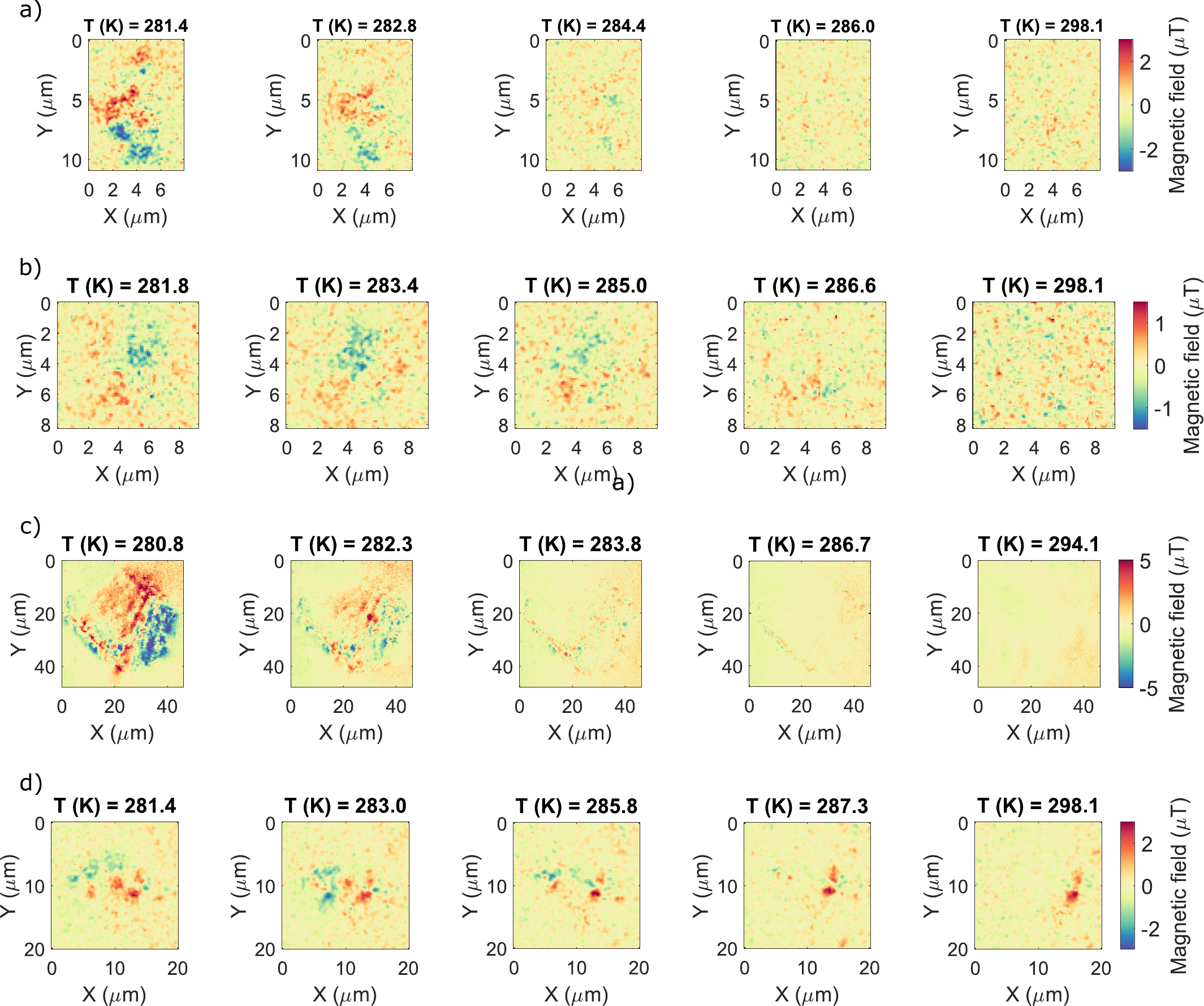}
  \caption{Magnetic field maps for Particles 2, 3, 4, and 5 ((a), (b), (c), (d) respectively) at selected temperatures. The field maps show the projection of the particle's magnetic field along the NV [1 1 1] axis for Particles 2 and 5 and the NV [-1 -1 1] axis for Particles 3 and 4 at each temperature. Note that the spatial and magnetic field scales are different between particles due to differences in size and signal strength. Particle 5 shows both a temperature-dependent signal and a spatially localized temperature-independent signal on the right side of the magnetic field map. The localized, temperature-independent magnetic field signal resembles a point dipole, consistent with a localized ferromagnetic impurity, while at low temperature an additional, more complicated canted antiferromagnetic signal appears originating from the rest of the particle.}\label{fig:temperature_maps_supp}
\end{figure}

\begin{figure}[htb]
  \centering
  \includegraphics[width=0.7\linewidth]{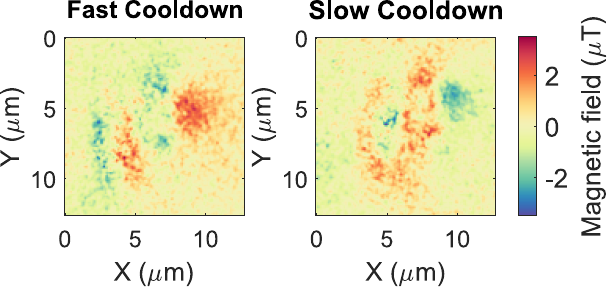}
  \caption{Magnetic field maps projected along the NV [1 -1 -1] axis of Particle 1 during a fast cooldown (left) and a slow cooldown (right). In the fast cooldown, the temperature setpoint was changed suddenly from 298 K to 278 K, equating to a diamond temperature of 282.9\,K, so the sample cooled at the maximum rate achievable with our setup, approximately 10 K/min. In the slow cooldown, the setpoint was ramped from 298 K to 278 K, resulting in a cooled diamond temperature of 282.7\,K, in 0.033 K steps over 20 min. Across four repetitions of this comparison, we observed no qualitative change in the number of domains formed between the two cooling methods, and no significant change in the magnitude of the magnetic signal: $\ev{\abs{B{_\text{fast}}}} - \ev{\abs{B{_\text{slow}}}} = 0.1\pm0.1\,\text{\textmu T}$ (mean $\pm$ SD across the four repetitions).}\label{fig:fast_vs_slow_supp}
\end{figure}

\begin{figure}[htb]
  \centering
  \includegraphics[width=0.9\linewidth]{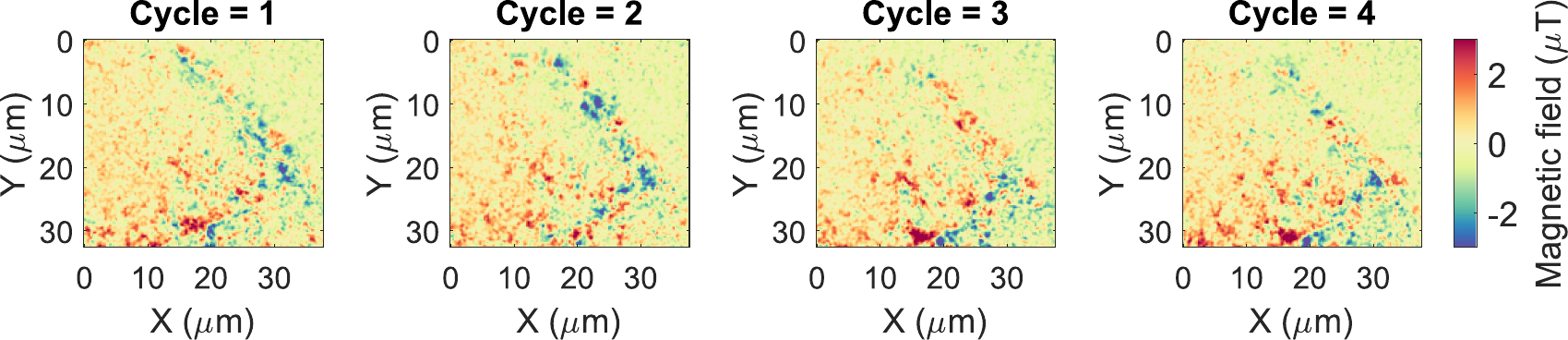}\\[1ex]
  \includegraphics[width=0.70344\linewidth]{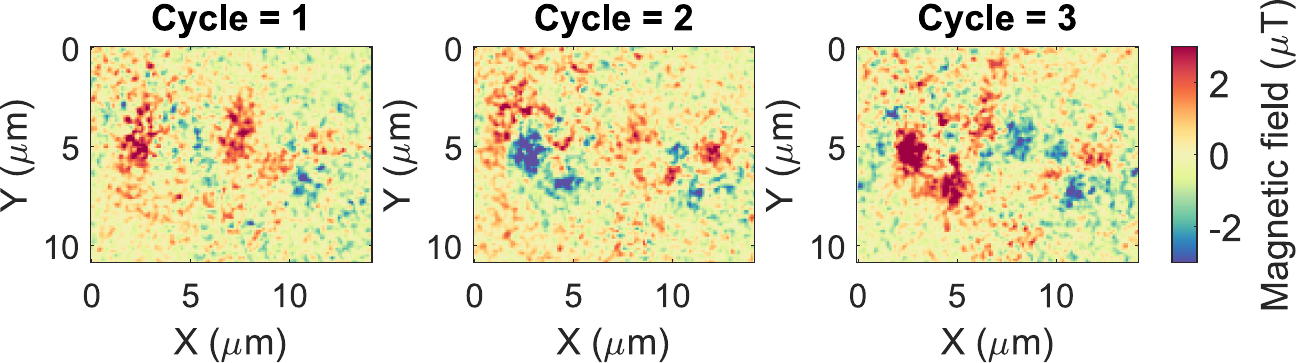}
  \caption{Temperature cycling data on Particle 4 (top) and Particle 5 (bottom) showing the magnetic field projection along the NV [-1 1 -1] axis. For each cycle, the particle was cooled below the Néel temperature, with diamond temperatures of $279.0\pm0.4$\,K and $282.2\pm0.4$\,K, respectively, then imaged. Between each cycle, the set temperature was raised to room temperature for a minimum of five minutes.}\label{fig:cycling_data_supp_p4p5}
\end{figure}

\begin{figure}[htb]
  \includegraphics[width=\linewidth]{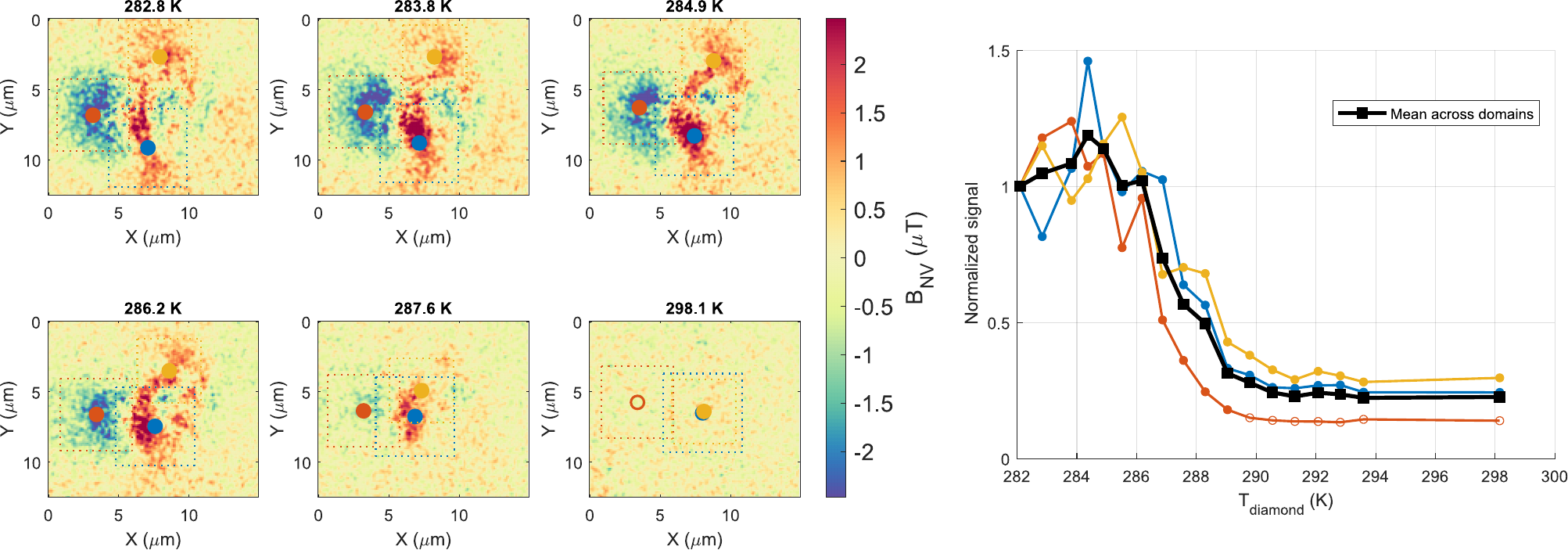}
  \caption{The magnetic field maps of the projection of Particle 1 along the NV [1 -1 -1] axis used to calculate the $\ev{\abs{B_{\text{domain}}}}$ curve in Fig. 2 of the main text. Individual domains with consistent  magnetic field sign were tracked over the course of a temperature scan. Due to sample drift and non-uniformity in domain shape vs. temperature, the signal from each domain is relatively noisy, though all show the same general trend in increasing magnetic signal strength with decreasing temperature. The mean signal from all tracked regions is shown in black in the lower plot.}\label{fig:domain_tracking_supp}
\end{figure}

\clearpage
\bibliography{LaCrO3_pub}